\documentclass[aps, showpacs, twocolumn]{revtex4}
\usepackage{graphicx}
\usepackage{dcolumn}
\usepackage{amsmath}
\usepackage{amssymb}
\usepackage{ulem}
\usepackage[colorlinks]{hyperref}
\hypersetup{citecolor=blue}
\usepackage{bm}
\usepackage{epstopdf}
\usepackage{amsthm}
\usepackage{float}
\usepackage{tabularx} 

\usepackage[caption=false]{subfig}

\usepackage{mathrsfs}
\usepackage{multirow, array, booktabs}

\def\prl#1#2#3{{ Phys. Rev. Lett.} {\bf #1}, #2 (#3)}
\def\prr#1#2#3{{ Phys. Rev. Res.} {\bf #1}, #2 (#3)}

\def\pre#1#2#3{Phys. Rev. E {\bf #1}, #2 (#3)}
\def\prx#1#2#3{Phys. Rev. X {\bf #1}, #2 (#3)}
\def\epl#1#2#3{{ Euro. Phys. Lett.} {\bf #1}, #2 (#3)}
\def\epjb#1#2#3{{ Euro. Phys. J. B} {\bf #1}, #2 (#3)}

\def\jsp#1#2#3{J. Stat. Phys. {\bf #1}, #2 (#3)}

\def\natphys#1#2#3{Nat. Phys. {\bf #1}, #2 (#3)}

\def\etl{$et~al.~$}

\def\la{\langle}
\def\ra{\rangle}

\def\beqr{\begin{eqnarray}}
\def\eqnr{\end{eqnarray}}
\def\beq{\begin{equation}}
\def\bc{\begin{center}}
\def\ec{\end{center}}
\def\eqn{\end{equation}\noindent}
\begin{document}

\title{Fitness Fluctuations and Correlation Time Scaling in the Barycentric Bak-Sneppen Model}

\author{Abdul Quadir}
\email{abdulq2013@gmail.com}
\author{Haider Hasan Jafri}
\email{haiderjaf@gmail.com}
\affiliation{Department of Physics, Aligarh Muslim University, Aligarh, 202 002, India}

\begin{abstract}
We consider the barycentric version of the Bak–Sneppen model, a one-dimensional self-organized critical model that describes generalized Keynesian beauty contests with a local interaction rule. We numerically investigate the power spectral density of the fitness variable and correlation time. Through data collapse for both variables, we estimate the critical exponents. For global and local fitness variables, the power spectral density exhibits $1/f^\alpha$ with $ 0 < \alpha < 2 $, indicative of long-range correlations. 
We also investigate the cover time, defined as the duration required for the extinction or mutation of species across the entire system in the critical state of the barycentric BS model. Using finite-size scaling and extreme value theory, we analyze the statistical properties of the cover time. Our results show power-law scaling with system size for the mean, variance, mode, and peak probability. Furthermore, the cumulative probability distribution exhibits data collapse, and the associated scaling function is well described by the generalized extreme value density, closely approximating the Gumbel family. 

\end{abstract}

\maketitle

\section{Introduction}

Self-organized criticality (SOC), introduced by Bak, Tang, and Wiesenfeld (BTW)~\cite{Bak_1987, Tang_1988, Bak_book1996}, provides a framework for understanding the $1/f^\alpha$ noise observed in non-equilibrium natural systems. SOC systems naturally evolve into a critical state where the system responds instantly to small perturbations, resulting in avalanches of random sizes. It was observed that the avalanche size and duration lack a characteristic scale that gives rise to power-law distributions. Likewise, temporal noise can exhibit low-frequency $1/f^\alpha$ behavior in its power spectral density, with the spectral exponent $\alpha$ typically ranging from $0$ (white noise) to $2$ (Brownian noise). Examples vary, from sandpile~\cite{Naveen_2022, Yadav_2012}, seismic activity~\cite{Gutenberg_1944, Omori_1895} to the biological systems like DNA sequences~\cite{Voss_1992}.  However, this scaling behavior disappears when the system is driven at a high external rate. SOC phenomena appear in a wide range of systems, including sandpile~\cite{Dhar_1989, Manna_1990, Yadav_2022, Zhang_1989, Sy_2024}, neuronal activities~\cite{Das_2019, Quadir_2024, Sole_2021} and the model of biological evolution~\cite {Bak_1993, Singh_2023, Boer_1994}.

The BS model, proposed by P. Bak and K. Sneppen is a paradigmatic example of SOC that encapsulates key aspects of coevolutionary dynamics and extremal selection~\cite{Bak_1993}. Initially developed to model biological evolution, it has since been applied to a wide range of complex systems exhibiting critical behavior. The model consists of $N$ species arranged in a ring topology. Each species is assigned a fitness value $f_i$ chosen randomly from the uniform distribution of interval $[0,1]$. Mutations can occur by identifying the species with the lowest fitness $f_{min}$, and replacing it, along with its interacting neighbors, with new random values drawn uniformly from the interval $[0,1]$. Through repeated iterations of this process, the system self-organizes into a stationary critical state, where each species has a fitness value above a threshold fitness $f_c \sim 0.667$~\cite{Paczuski_1996}. The model dynamics are a two-step process- mutation and evolution.
Recent studies suggest that the fitness noise follows $1/f^\alpha$ behavior with a spectral exponent $\alpha=1.2$ for a one-dimensional lattice and $\alpha=2$ for the mean field version of the BS model~\cite{Singh_2023}. Beyond evolutionary biology, the BS model has been employed in various domains, including ecosystem stability~\cite{Li_2000}, financial market dynamics~\cite{Yamano_2001}, and information spreading in complex networks~\cite{Valleriani_1999}. Several extensions, such as higher-dimensional versions~\cite{Chhimpa_2024i} and modified update rules~\cite{Singh_2023, Chhimpa_2025i} have been explored while preserving the model's self-organized critical behavior.

Our interest is in the one-dimensional BS model based on the phenomenon of conformity, referred to here as the local barycentric BS model~\cite{Kennerberg_2021}. This model combines the classical Bak–Sneppen (BS) model~\cite{Bak_1993} with a formalization of Jante's law, originally introduced to describe social norms in Scandinavian countries that discourage individuals from standing out \cite{Sandemose_1936, Grinfeld_2015, Kennerberg_2018}. In interacting particle systems, Jante’s law has been modeled as the ``Keynesian beauty contest process,'' which captures dynamics of conformity and competition among agents.  
This model reflects a sociological analogy in which species strongly differing from their local environment (``least conformist" sites) are selectively disadvantaged, while those closer to the neighborhood mean persist~\cite{Sandemose_1936}.
While the barycentric Bak–Sneppen model preserves the extremal-dynamics framework of the original BS model, it introduces a fundamentally different update rule: instead of replacing the site of minimum fitness and its neighbors by independent random numbers, we perform a barycentric (conformity-based) averaging with neighboring fitness values. This modification induces local correlations and a smooth redistribution of fitness, leading to broader temporal correlations and slower relaxation. This selection mechanism reflects the conformity-driven interactions and provides a tractable framework for exploring self-organized criticality~\cite{Kennerberg_2021, Grinfeld_2015}. Like the classical BS model, each species possesses a fitness value, but the key dynamics here are determined by relative deviations rather than absolute fitness. A species with the most deviated fitness is considered a failure, and extinction occurs together with its interacting neighbors. Competition between ordered (successful) and disordered (failed) species drives criticality in the system. Despite the simplicity of the local barycentric BS model, it exhibits robust behavior: the least conformist site in space-time shows fractal structure, and its trajectory follows a Lévy flight pattern with jump sizes distributed according to a power law~\cite{Singh_2024}.
Quantitatively, we find that the change in the spectral exponent and finite-size scaling exponent change, indicating stronger long-range memory. 
It shows that the power spectrum remains consistent up to a certain low-frequency limit, below which the spectrum becomes uncorrelated~\cite{Singh_2024}. 

Despite the simplicity of the barycentric BS model, it exhibits robust behavior: the least conformist site in space-time shows fractal structure, and its trajectory follows a Lévy flight pattern~\cite{Singh_2024}. In this context, the time required for the extinction or mutation of all the species is termed as cover time. We emphasize that the numerical value of the cover time exponent is equal to the avalanche dimention~\cite{Paczuski_1996, Boettcher_2000} which reflect the interconnected nature of space-time correlations. Recent studies suggested that the cover time follows the Gumbel distribution~\cite{Chupeau_2015, Barkai_2015}. This cover time quantifies the typical distance over which a system exhibits nontrivial correlation~\cite{Chhimpa_2025}.
In the thermodynamic limit, the critical Ising model exhibits a diverging cover length and an algebraically decaying correlation function. Palmieri and Jensen recently studied critical models where the instantaneous, time-fluctuating cover length acts as a stochastic variable. Its average defines the cover length, while its distribution reveals deeper aspects of critical behavior~\cite{Palmieri_2020}. These results show that the barycentric formulation constitutes a distinct subclass within the BS universality family-still self-organized and critical, but governed by conformity-driven, correlated updates that generate a broader $1/f^\alpha$ spectrum and modified finite-size scaling.

This paper aims to reveal a subtle understanding of the $1/f^{\alpha}$ noise and cover time in the model.  In a class of SOC models, the space-time correlations can be studied with the help of finite-size scaling (FSS).  The FSS is useful to get the scaling functions and the critical exponents. Our analysis reveals $1/f^{\alpha}$ noise for different fluctuations in the barycentric BS model. The cutoff frequency is found to vary as $f_0 \sim L^{-\lambda}, ~\lambda=2.45$ with the system size $L$. The two independent spectral exponents $\alpha$ and $\lambda$ characterize the spectral properties. The scaling relation between $\alpha$ and avalanche dimension $D$ is given by $\alpha = 1- 1/D$~\cite{Paczuski_1996}. For local activity, we show $\lambda=D$, which indicates the existence of a single independent critical exponent. The cover time can be understood in many alternative ways. To explore the cover in the barycentric BS model, we first examine the system size dependence of various statistical quantities. Our scaling function argues that the critical exponents for different statistical quantities are the same. The data collapse curve for the cumulative probability distribution (CDF) of scaled correlation time fits reasonably well with the GEV distribution. We numerically calculate the critical exponent $\lambda$ using different methods and found the same values in all the methods.

The paper is organized as follows. Section~\ref{sec-model} describes the barycentric BS model and its different variants based on the interaction rules. In Sec.~\ref{sec-results}, we describe the numerical and analytical results for the power spectra of local fitness, global fitness (sum of all the species) fluctuations and statistical aspects of correlation time. We also present the data collapse and fitting with the generalized extreme value (GEV) theory. Finally, we conclude with a summary and discussion in Sec.~\ref{sec-conclusion}.

\section{Model}~\label{sec-model}

Consider a lattice having size $L \geq 3$ with periodic boundary conditions. Each lattice point is assigned a random fitness $\xi_i$ chosen randomly from a uniform distribution in the interval $[0,1]$. The lattice site that deviates the most from the average fitness of its neighbors will be replaced, along with its neighbors, by a new randomly chosen fitness $\xi^{new}_i \in [0,1]$. For a given lattice having size $L\geq 3$, let $i = \{1,2,3,.....L \}$ be a set of lattice points on a periodic lattice (or ring). At time $t$, each lattice $i$ has a certain `fitness' $\xi_i(t) \in {R}$. Thus, the deviation from  average fitness is defined as

\beq 
\mathcal{D}_i(t) = \left\lvert \xi_i(t) - \dfrac{\xi_{i+1}(t) +\xi_{i-1}(t)}{2} \right\rvert 
\eqn  
Since the lattice is  periodic, $\xi_{L+1} = \xi_1$ and $\xi_0 = \xi_L$. The lattice site $ j = \max_{i \in L} \{ \mathcal{D}_i(t) \}$ represents the most deviated lattice site at time $t$ and is termed as the ``least conformist site''. 

\begin{figure}[htb]
    \centering
    \includegraphics[scale=0.4]{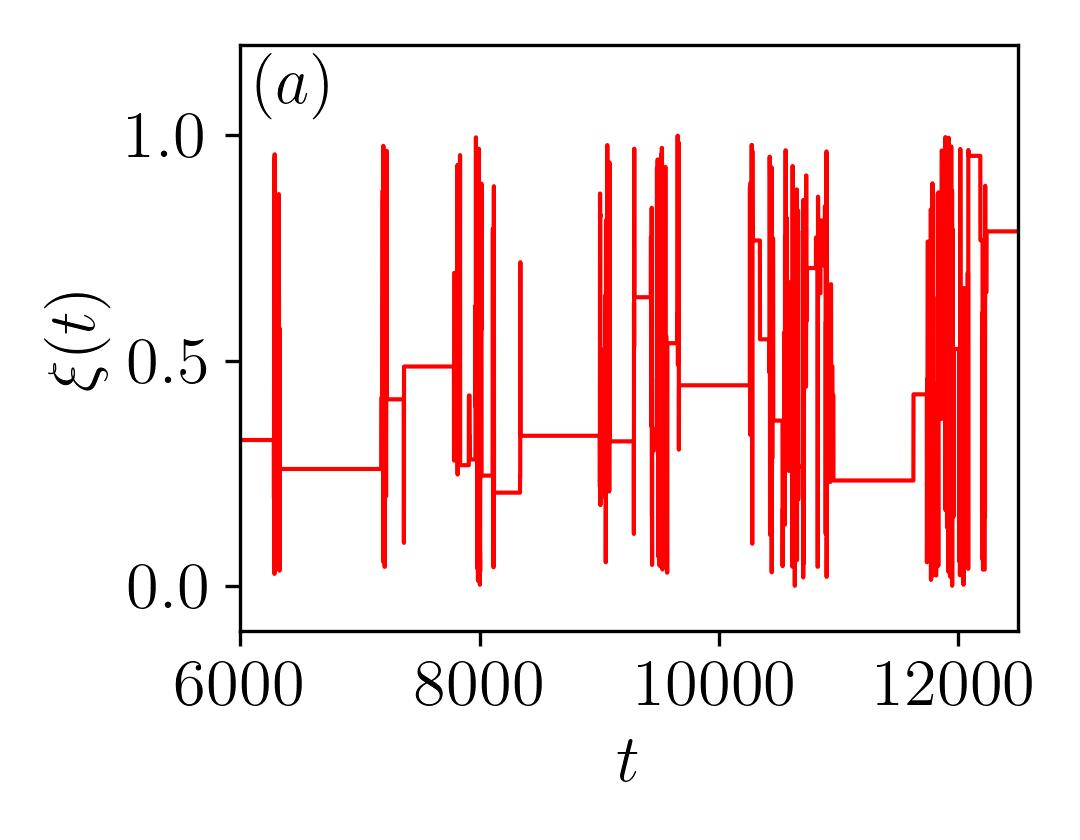}
    \includegraphics[scale=0.41]{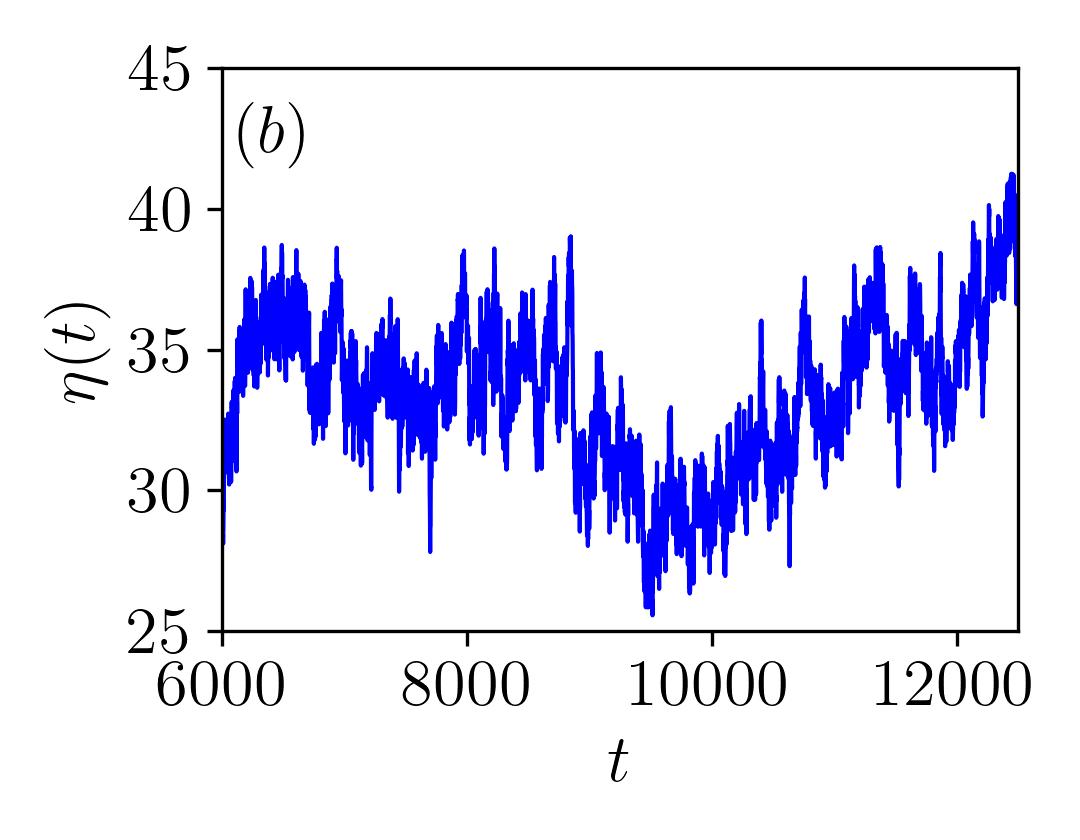}
    \caption{ (a) Time series of the local fluctuation $\xi_i(t)$ of the least conformist site $i=2^2$ for a system having size $L=2^6$. It shows intermittency in burst activities with punctuated equilibrium features. In (b), we plot the time series of the  global fitness signal $\eta(t)$ for system size $L=2^6$.}~\label{Fig-ts}
\end{figure}

Our interest is in the fitness fluctuations over time. Therefore, global fitness in terms of the local fitness is defined as $\eta(t) = \sum_{i=1}^{L} \xi_i(t)$ with average fluctuation, given by $\bar{\eta}(t) = {\eta}(t)/L$ ~\cite{Li_2000, Singh_2023, Chhimpa_2024, Chhimpa_2024i}. For the barycentric BS model, Fig.~\ref{Fig-ts} shows the typical time series of (a) the local fitness $\xi(t)$ and (b) global fitness $\eta(t)$.

\subsection{Variants of the Model}
Different variants of the barycentric BS model are introduced to test the robustness of self-organized criticality (SOC) against changes in the interaction rules. 
The original barycentric BS model (model~1) updates the least conformist site together with its two nearest neighbors. 
By modifying this rule of interactions, one can assess whether the SOC behavior is a generic property of the model or if it relies on specific update rules. 
Thus, the variants serve as controlled perturbations to the original dynamics, allowing us to investigate universality, crossover phenomena, and the sensitivity of scaling exponents to changes in interaction structure~\cite{Singh_2024, Singh_2023, Chhimpa_2024i}. In the following, we briefly describe the rules for these models.
\begin{enumerate}
    \item Model 1: Two nearest neighbor interaction.
    \item Model A: only one nearest neighbor interaction.
    \item Model B: Only one random neighbor interaction from left or right, chosen with equal probability. 
    \item Model C: The random neighbor interaction includes two sites chosen randomly with equal probability among the remaining $L-1$ sites.
\end{enumerate}

\section{Results}~\label{sec-results}
In this section, we study the temporal behavior of global and local fluctuation as well as the correlation time for the barycentric BS model with its variants. The PSD is useful to understand the temporal behavior of the fitness fluctuations.

\subsection{The fitness fluctuations and power spectral analysis}~\label{sec-fitness}

In this work, we study the local $\xi(t)$ and global $\eta(t)$ fitness fluctuations for the barycentric BS model and its variants by calculating the power spectral density (PSD). The PSD is useful to understand the temporal behaviour of noisy signal. The PSD is defined as the Fourier transformation of the two-time autocorrelation functions. To evaluate the PSD, we first calculate $\Tilde{x}(t)$, the Fourier transformation of the noisy signal $x(t) \in \{  \xi(t), \eta(t) \}$ where $t=1,2,...,N$ by using the standard fast Fourier transformation (FFT) algorithm. Thus, the PSD is given by
\beq 
\mathcal{S}(f) = \lim_{N \to \infty } \dfrac{1}{N} \la \mid \Tilde{x}(f) \mid^2 \ra
\eqn 
where $\la \cdot \ra$ is the ensemble average over $M$ different realizations of the fitness signal $x(t)$. The Fourier component of $x(t)$ is thus, defined as 
\beq  
\Tilde{x} (f=k/N) = \mathcal{F}(x(t)) = \dfrac{1}{\sqrt{N}} \sum_{t=0}^{N-1} x(t) \exp \left( -2\pi j \dfrac{k}{N}t \right),
\eqn 
Here, $\mathcal{F}(x)$ denotes the fast Fourier transform (FFT). For numerical results, we use Monte Carlo simulations to obtain the noise signal $x(t)$. The PSD is computed for signal length $N=2^{18}$ to $2^{20}$ after discarding $10^6$ transients, with ensemble averages performed over $M=10^4$ realizations for system sizes $L=2^4,2^5,2^6,2^7$.  

\begin{figure}[h]
    \centering
    \includegraphics[scale=0.45]{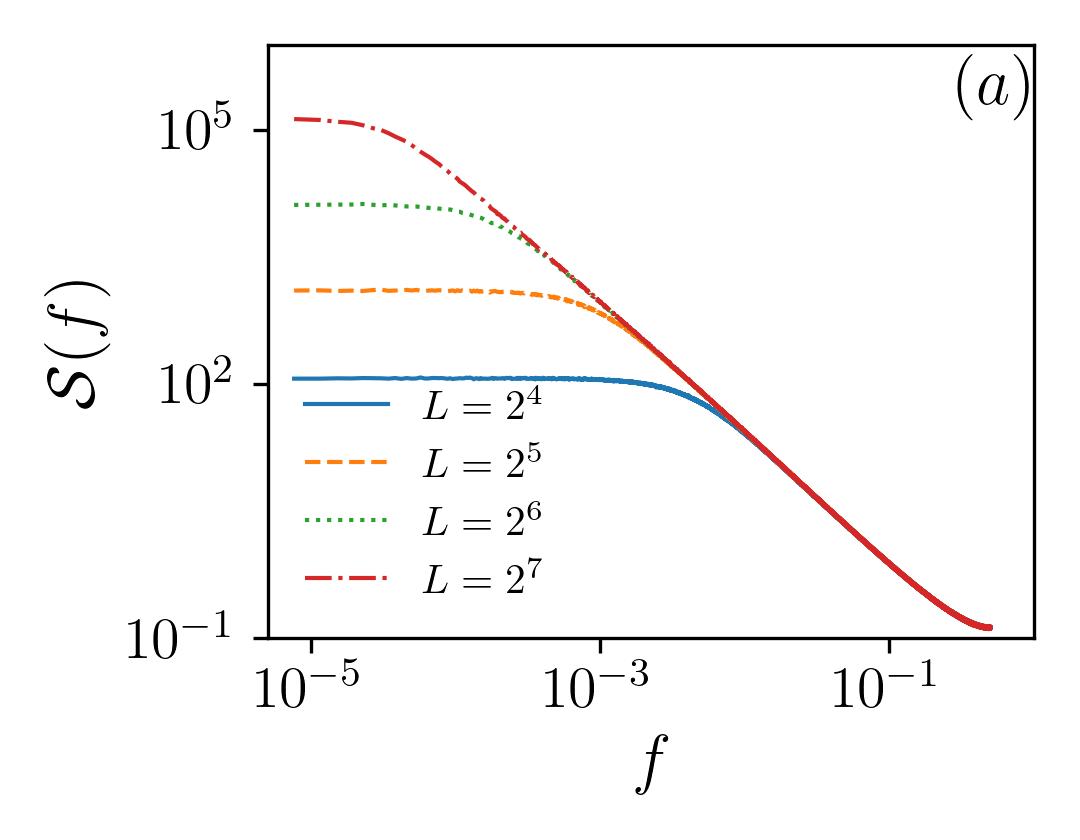}
    \includegraphics[scale=0.45]{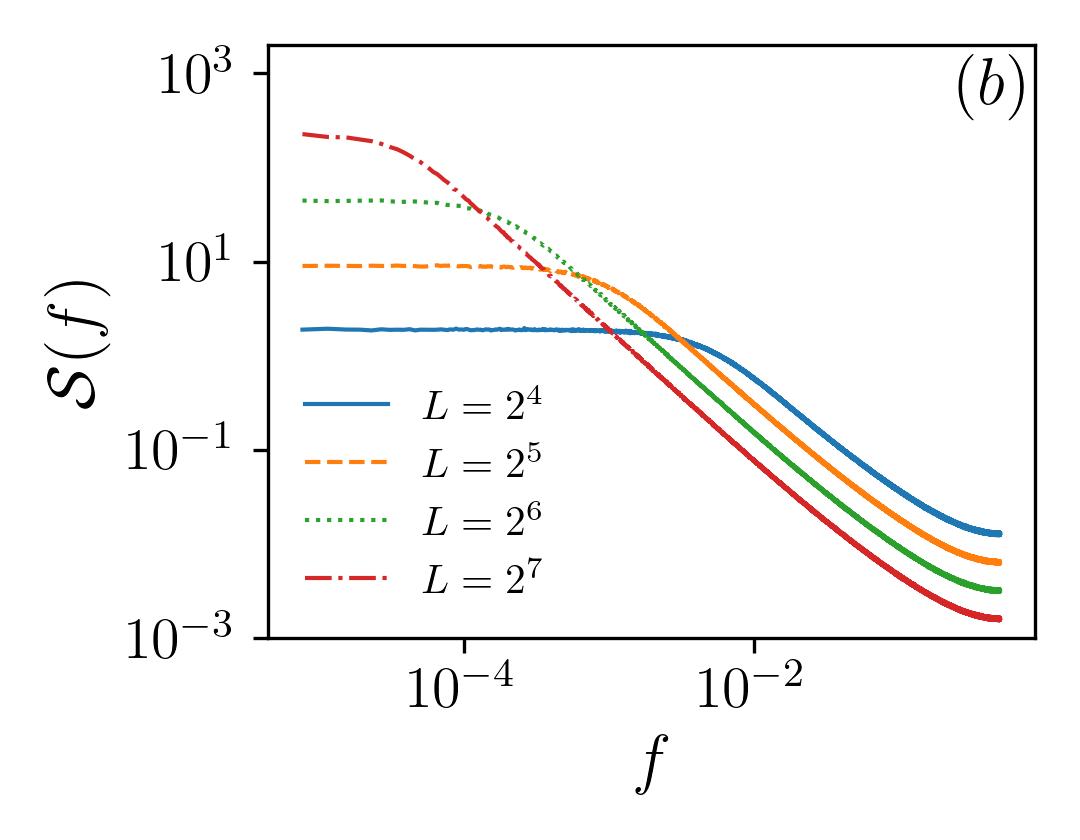} \\
    \includegraphics[scale=0.45]{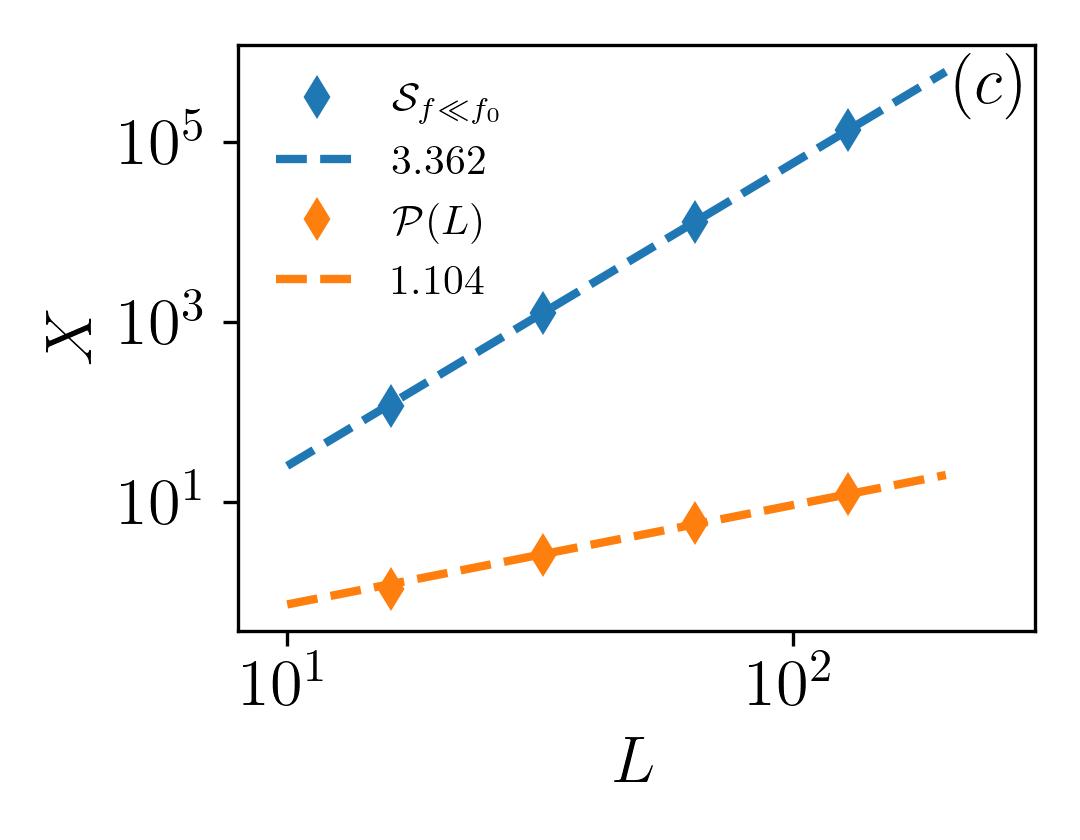} 
    \includegraphics[scale=0.45]{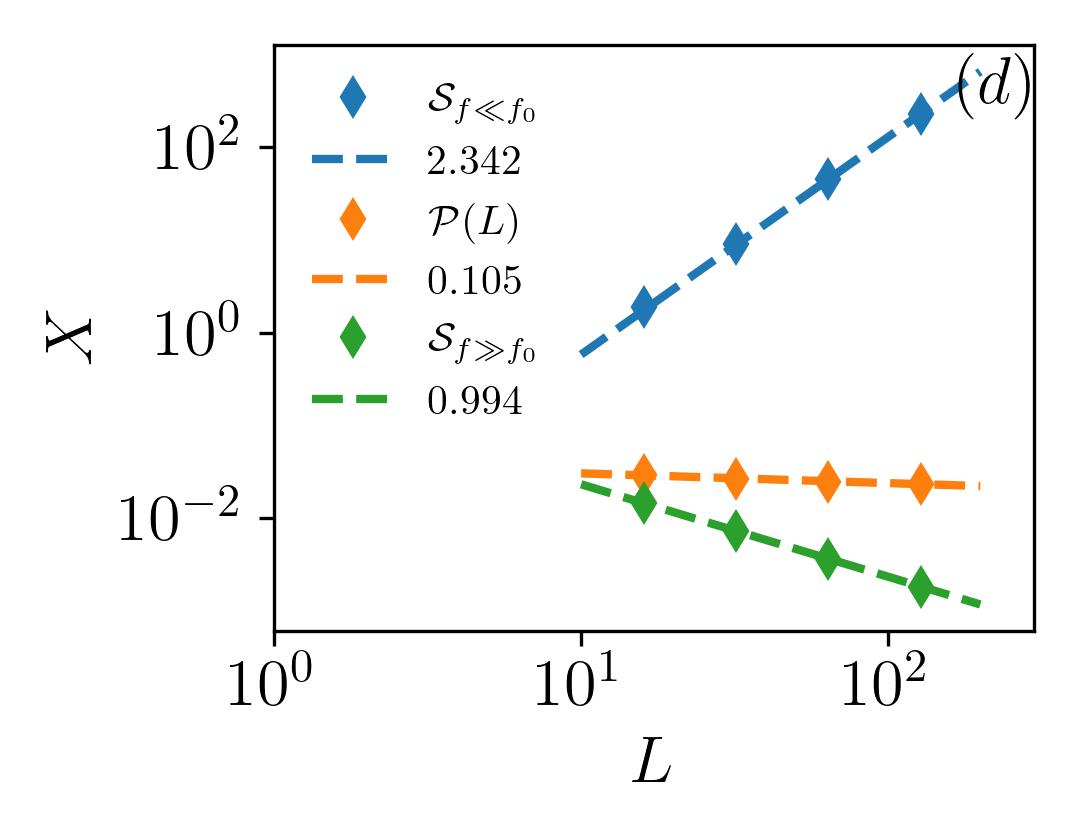} \\
    \includegraphics[scale=0.45]{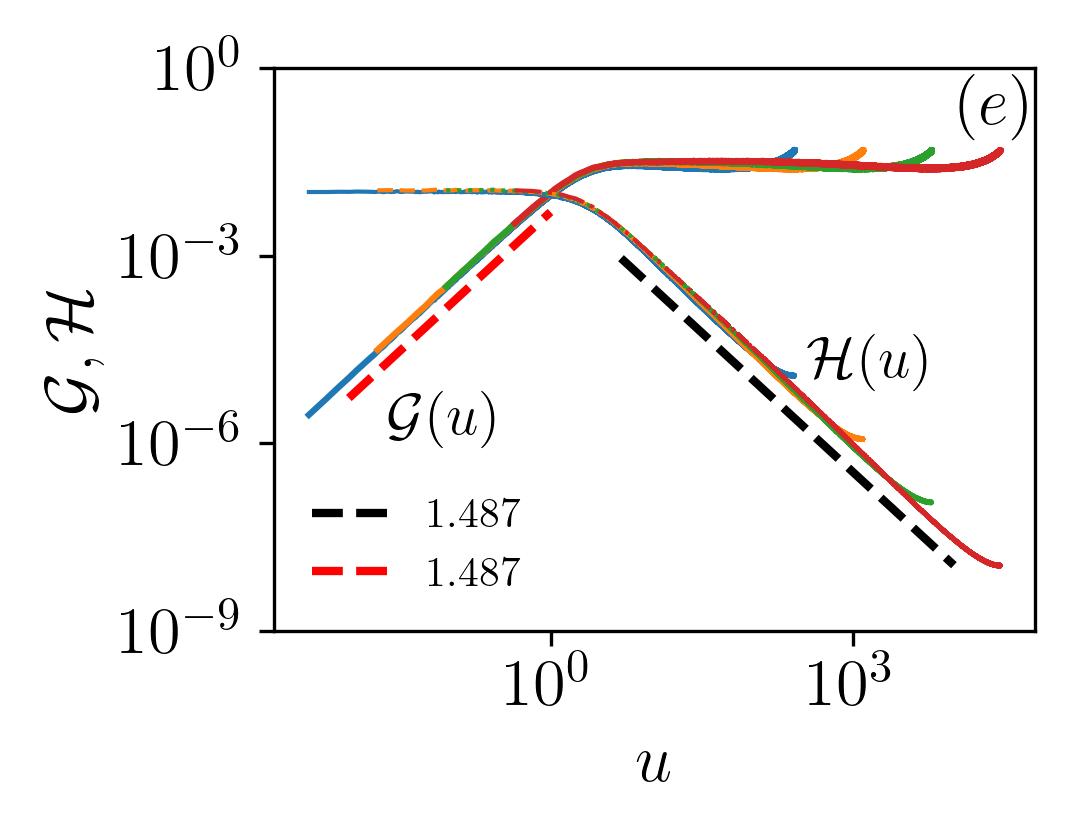}
    \includegraphics[scale=0.45]{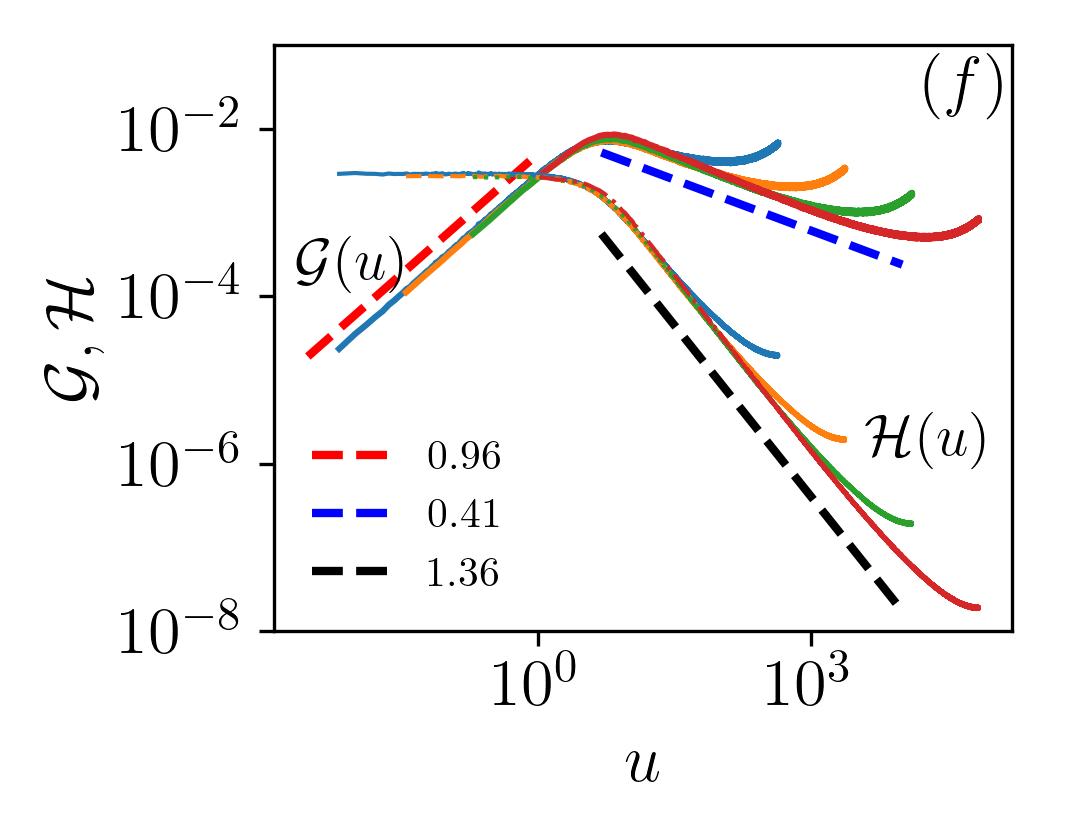}
    \caption{ {\bf For model~1}: Left panel -- {\bf (a)} The power spectra for global fitness fluctuations for system size $L=2^4, 2^5, 2^6, 2^7$; {\bf (c)} The power at a fixed frequency below and above the cut-off $f_0$ including the total power $\mathcal{P}(L)$, along with the best-fit curves. The estimated exponents are $a = 3.36$ and $c = 1.10$. {\bf (e)} The scaling functions for the global fitness power spectra.  Here, $b=0$ which implies $\lambda = a-c = 2.26$ and $\alpha = a/\lambda = 1.49$. Right panel: Same as left panel, but for local fitness fluctuations. Here, the estimated exponents are $a=2.34,~b=0.99,$ and $c=-0.11$ which imply $\lambda = 2.45$ and $\alpha=1.36$.}
    \label{Fig-Model1}
\end{figure}


\begin{figure}[htb]
    \centering
    \includegraphics[scale=0.45]{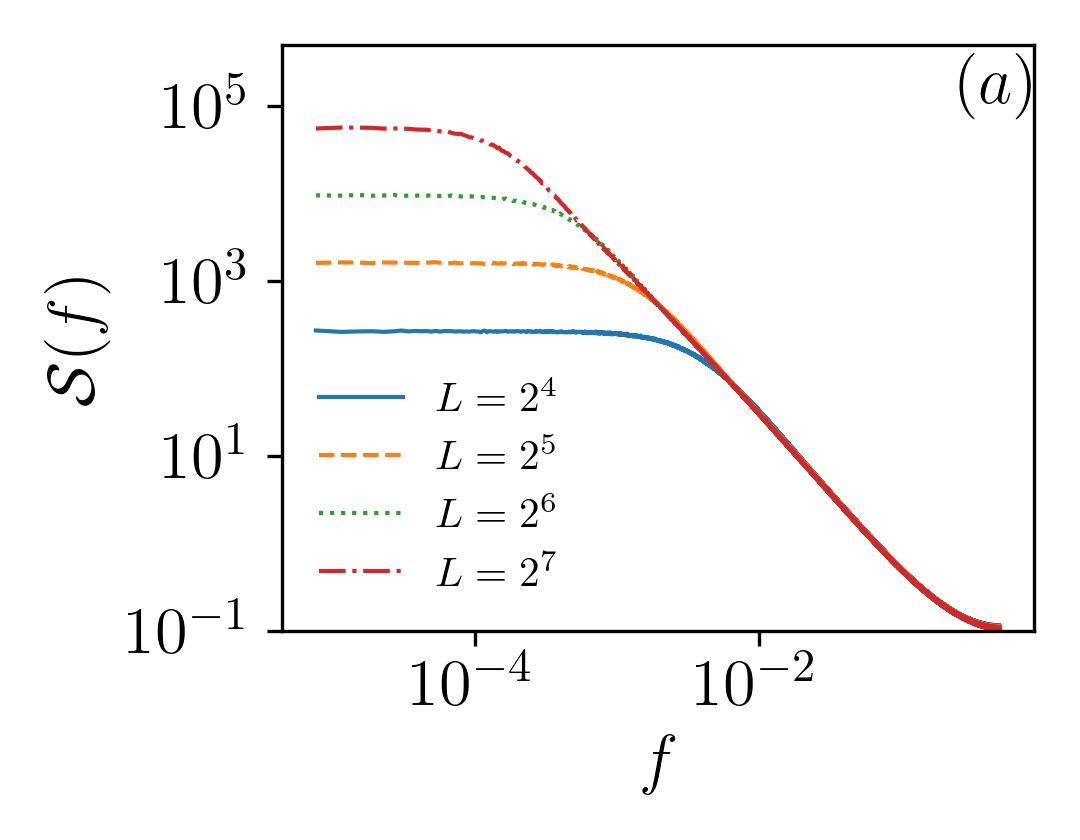}
    \includegraphics[scale=0.45]{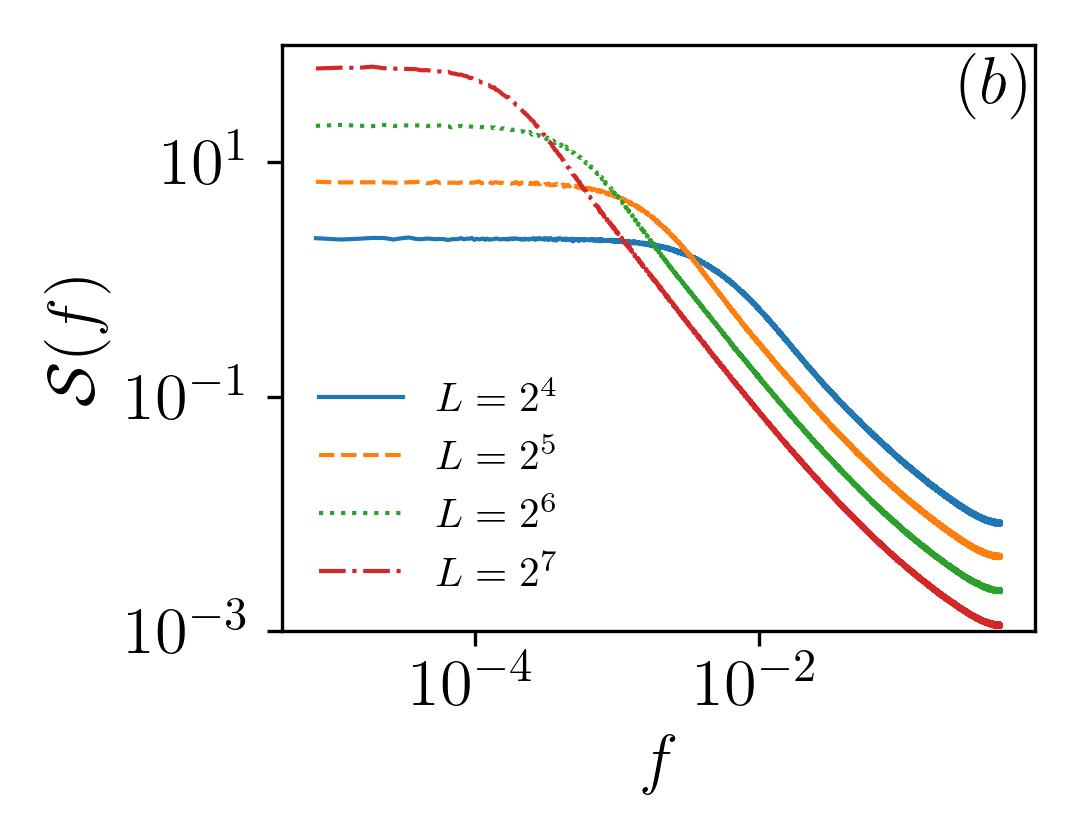} \\
    \includegraphics[scale=0.45]{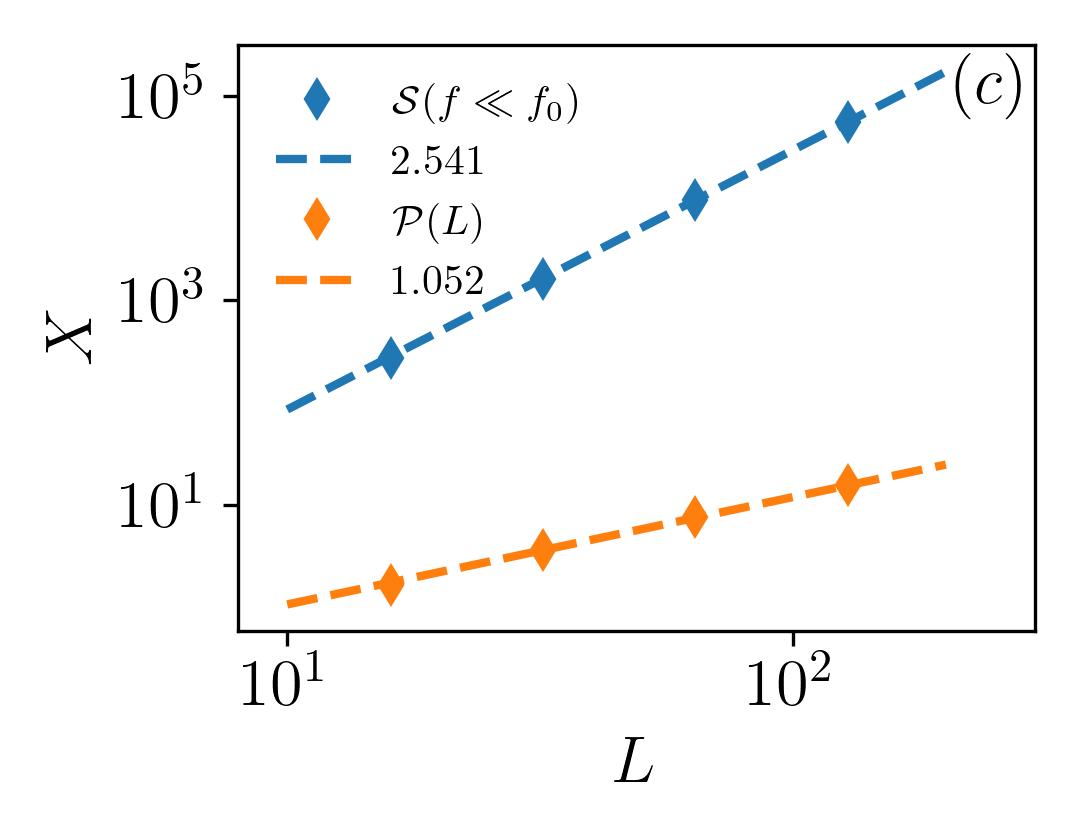}
    \includegraphics[scale=0.45]{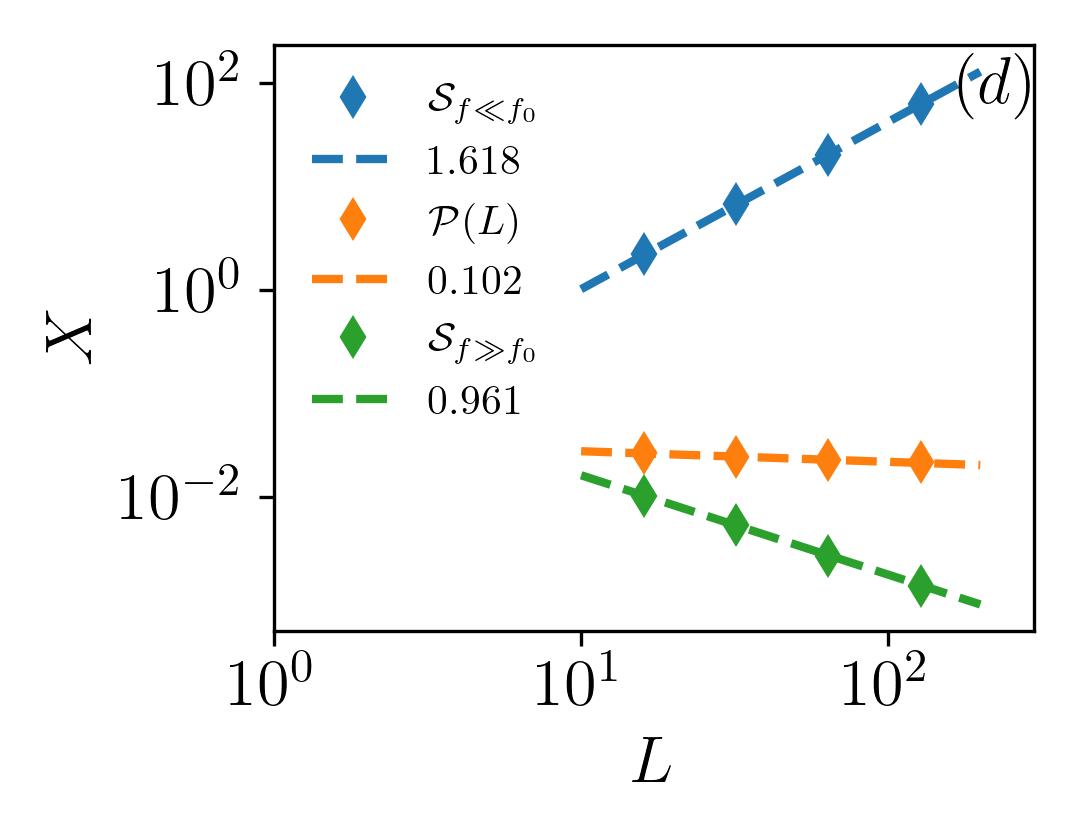} \\
    \includegraphics[scale=0.45]{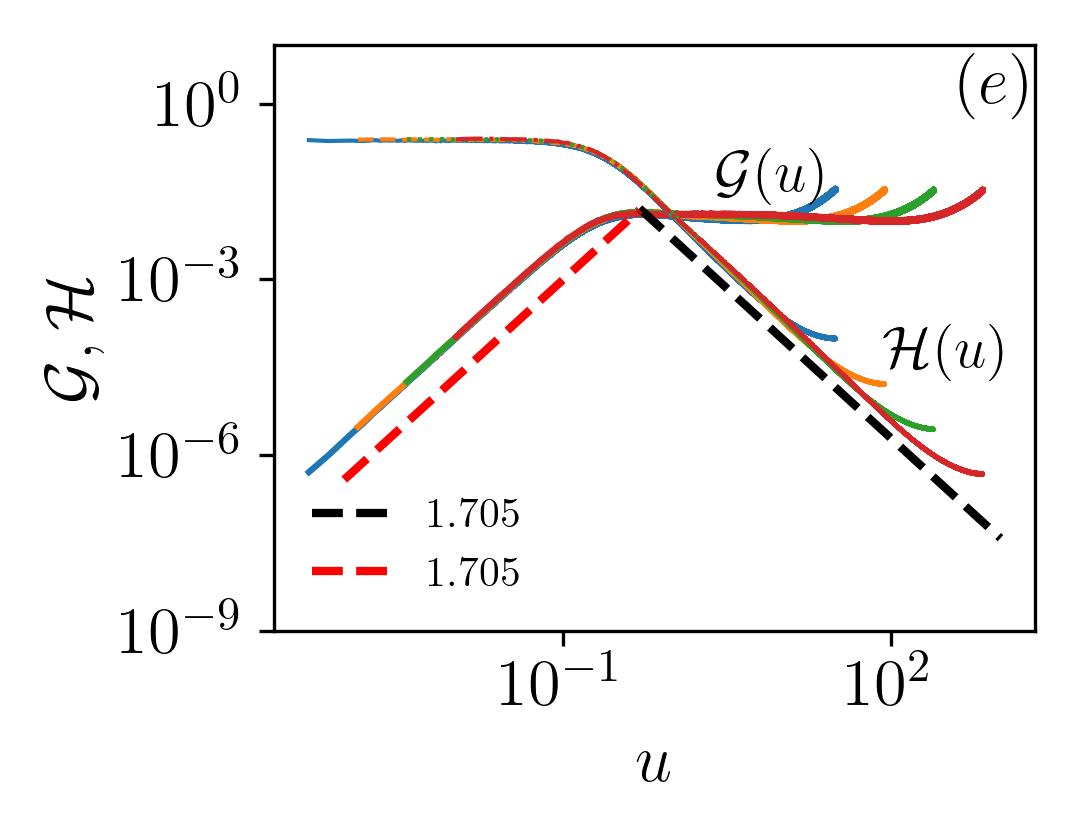}
    \includegraphics[scale=0.45]{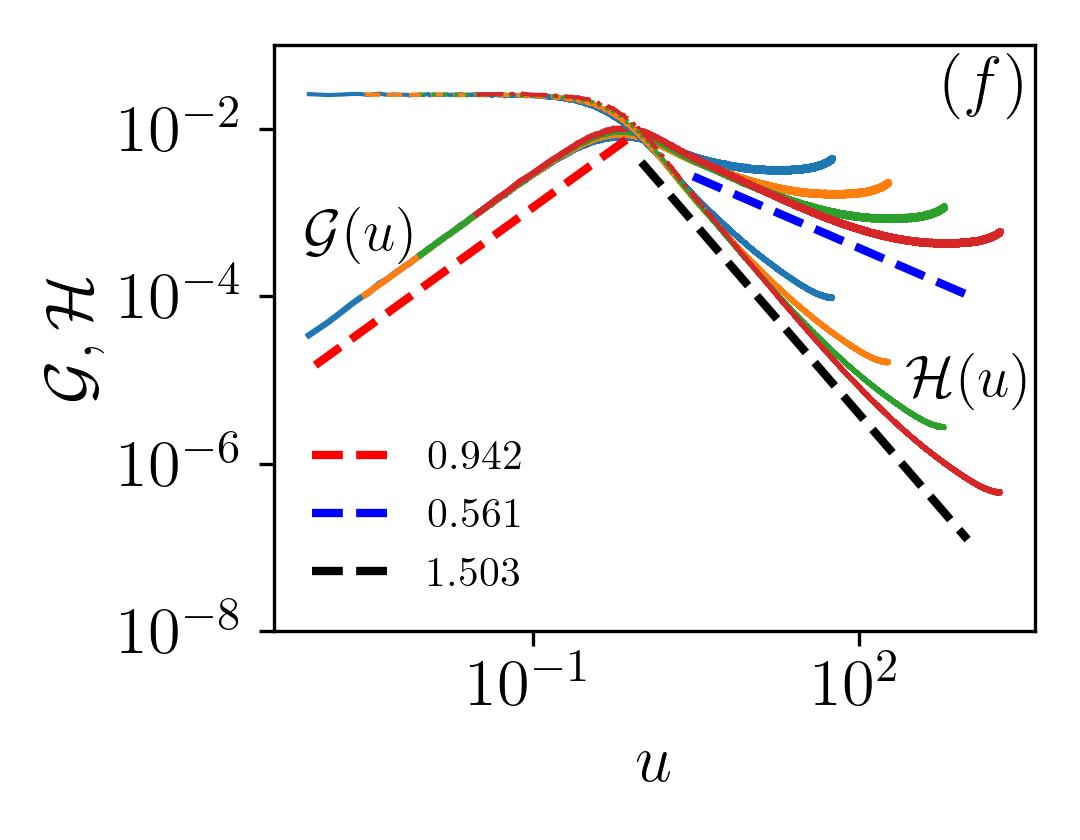}
    \caption{ {\bf For model~A}: Left panel -- {\bf (a)} The power spectra for global fitness fluctuations for system size $L=2^4, 2^5, 2^6, 2^7$; {\bf (c)} The power at a fixed frequency below and above the cut-off $f_0$ including the total power $\mathcal{P}(L)$, along with the best-fit curves. The estimated exponents are $a = 2.54$ and $c = 1.05$. {\bf (e)} The scaling functions for the global fitness power spectra.  Here, $b=0$ which implies $\lambda = a-c = 1.49$ and $\alpha = a/\lambda = 1.70$. Right panel: Same as the left panel, but for local fitness fluctuations. Here, the estimated exponents are $a=1.62,~b=0.96,$ and $c=-0.10$ which imply $\lambda = 1.72$ and $\alpha=(a+b)/\lambda = 1.50$.}~\label{Fig-ModelA}
\end{figure}


\begin{figure}[htb]
    \centering
    \includegraphics[scale=0.45]{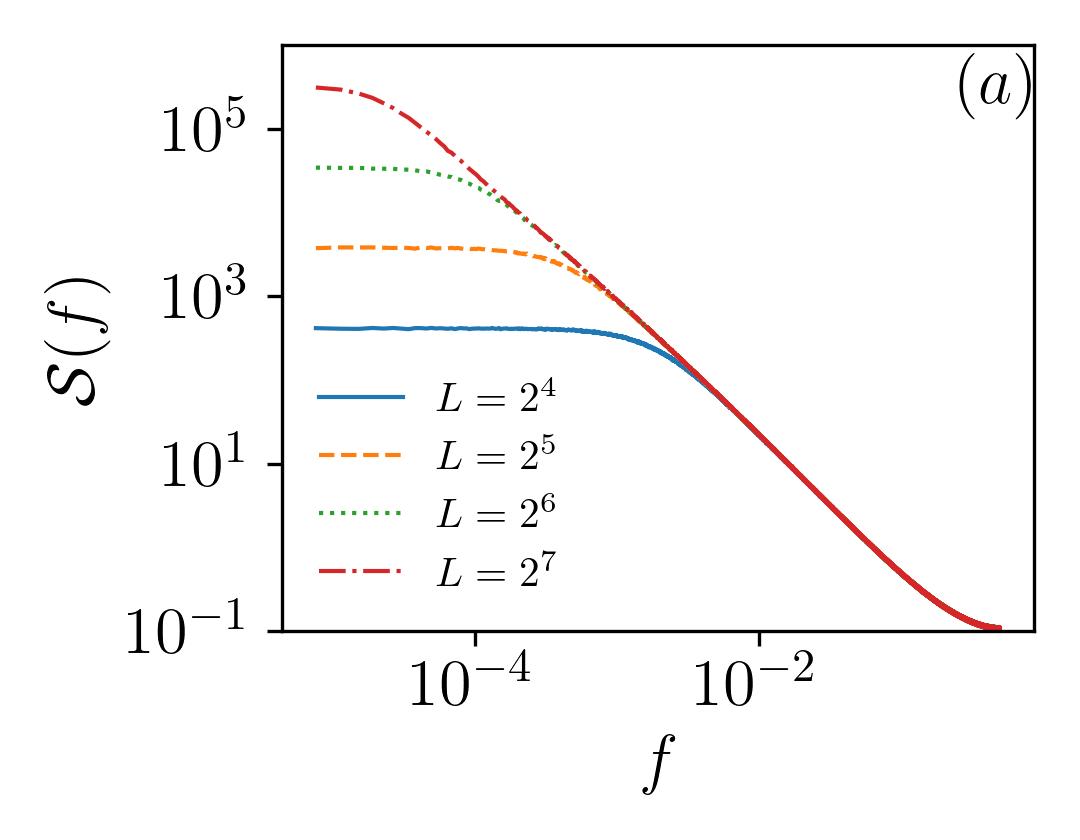}
    \includegraphics[scale=0.45]{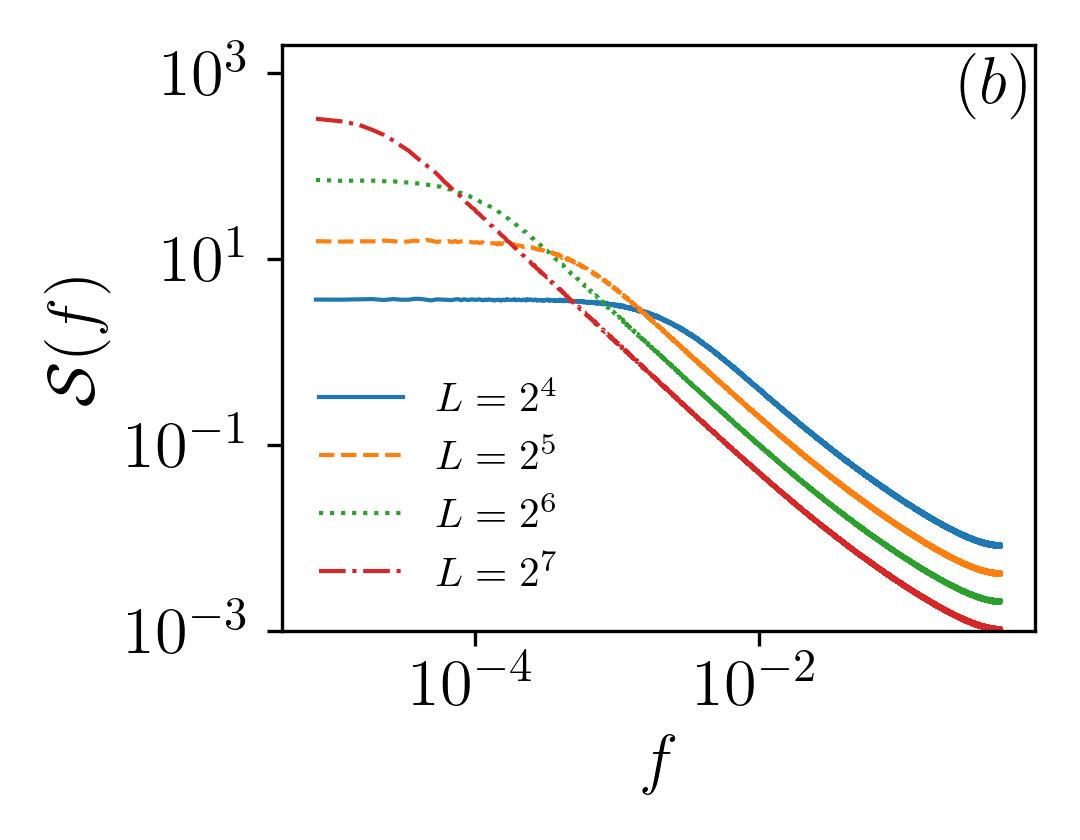} \\
    \includegraphics[scale=0.45]{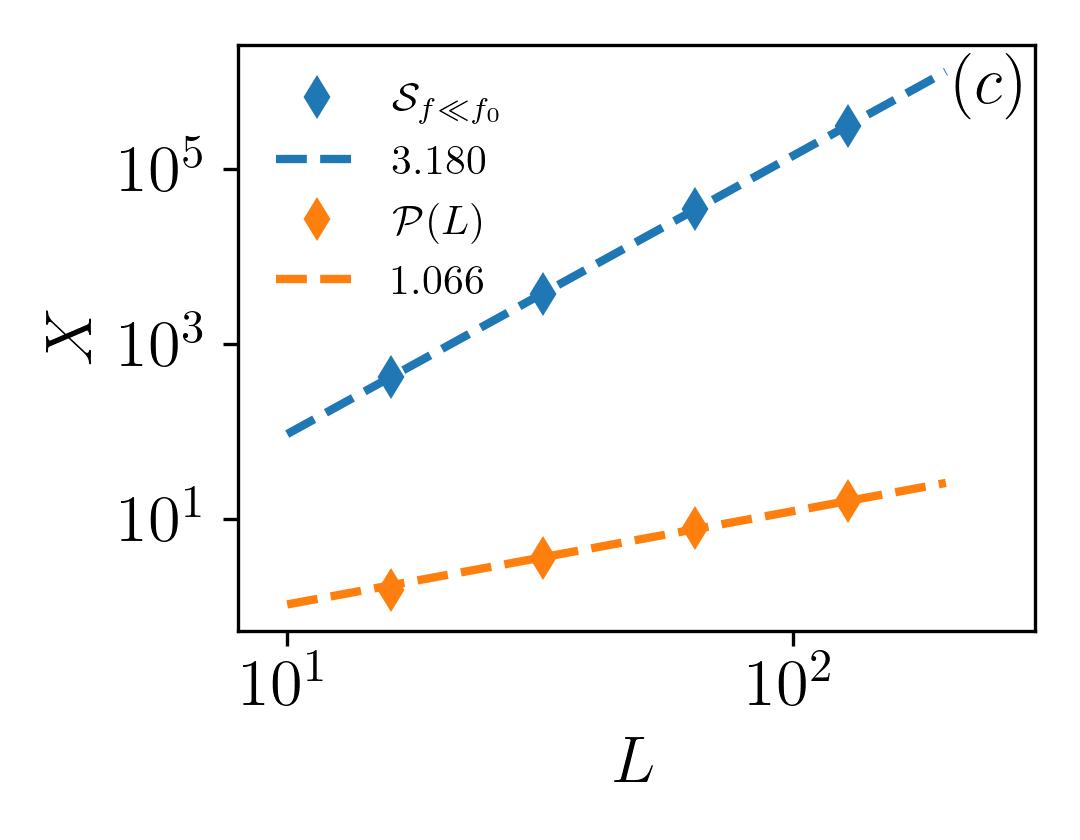}
    \includegraphics[scale=0.45]{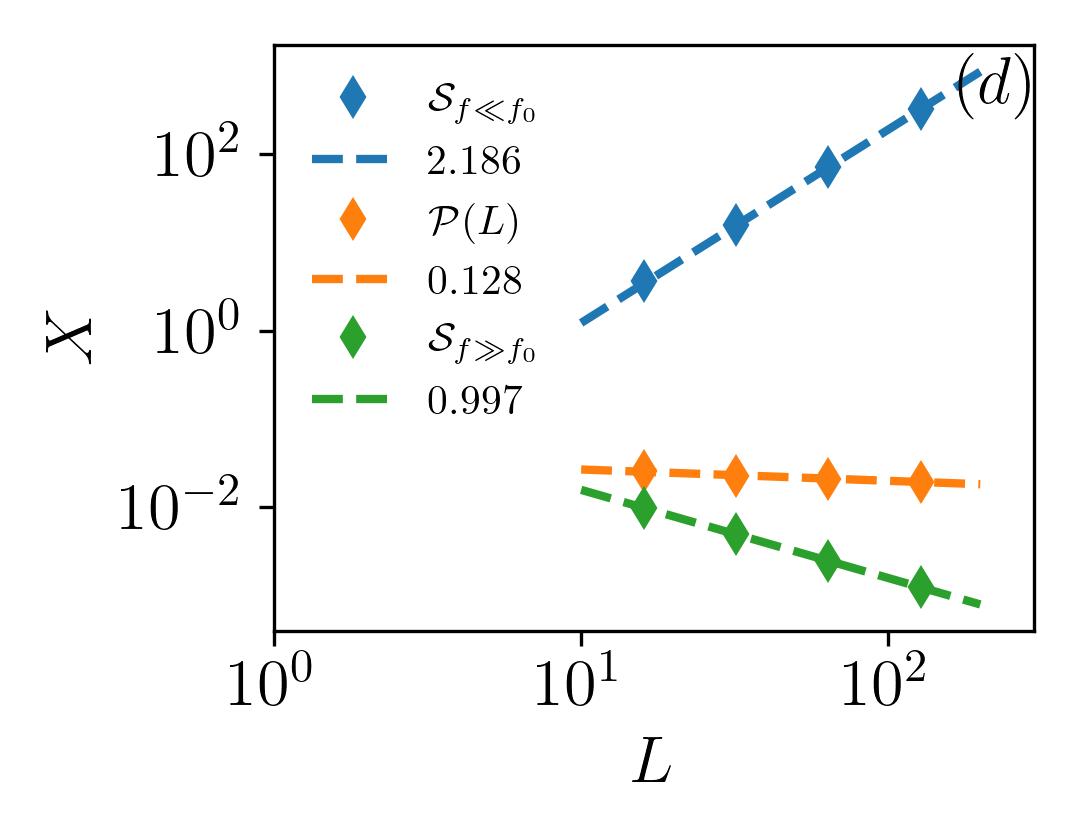} \\
    \includegraphics[scale=0.45]{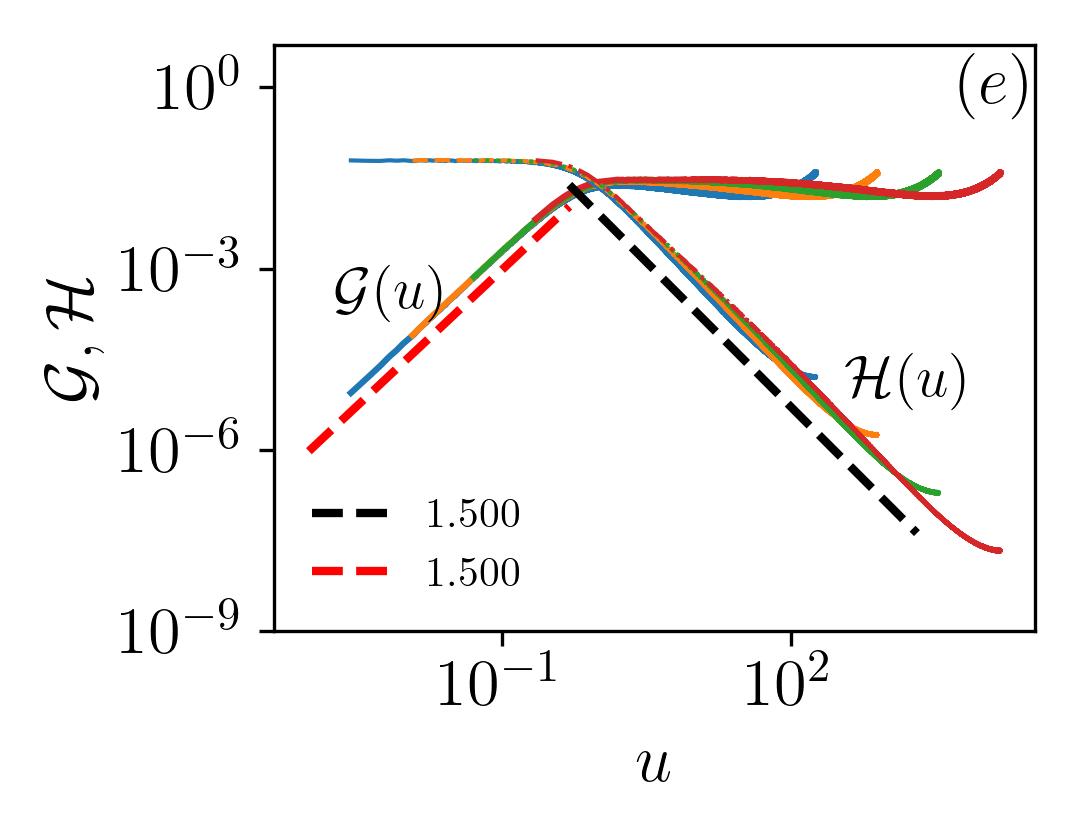} 
    \includegraphics[scale=0.45]{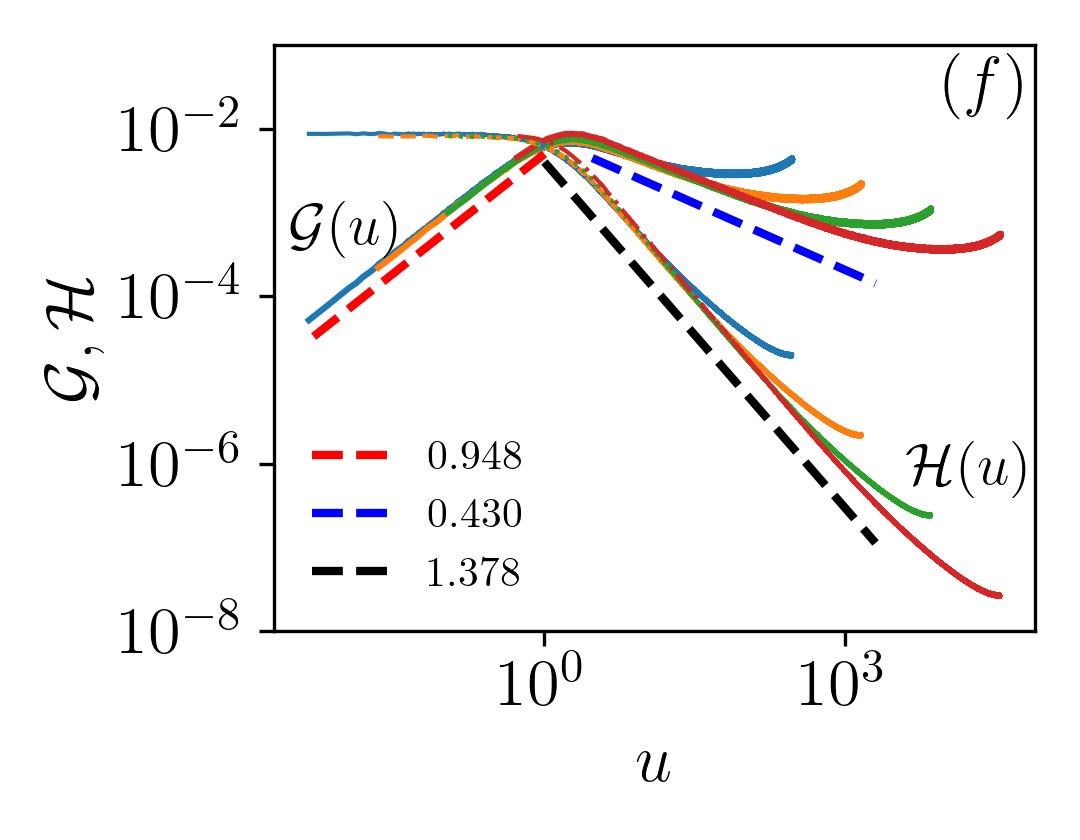} 
    \caption{ {\bf For model~B}: Left panel -- {\bf (a)} The power spectra for global fitness fluctuations for system size $L=2^4, 2^5, 2^6, 2^7$; {\bf (c)} The power at a fixed frequency below and above the cut-off $f_0$ including the total power $\mathcal{P}(L)$, along with the best-fit curves. The estimated exponents are $a = 3.18$ and $c = 1.07$. {\bf (e)} The scaling functions for the global fitness power spectra.  Here, $b=0$ which implies  $\lambda = a-c = 2.11$ and $\alpha = a/\lambda = 1.50$. Right panel: Same as the left panel, but for local fitness fluctuations. Here, the estimated exponents are $a=2.19,~b=1.00,$ and $c=-0.13$ which imply $\lambda = 2.32$ and $\alpha=(a+b)/\lambda =(a+b)/\lambda =1.38$.}~\label{Fig-ModelB}
\end{figure}


\begin{figure}[htb]
    \centering
    \includegraphics[scale=0.45]{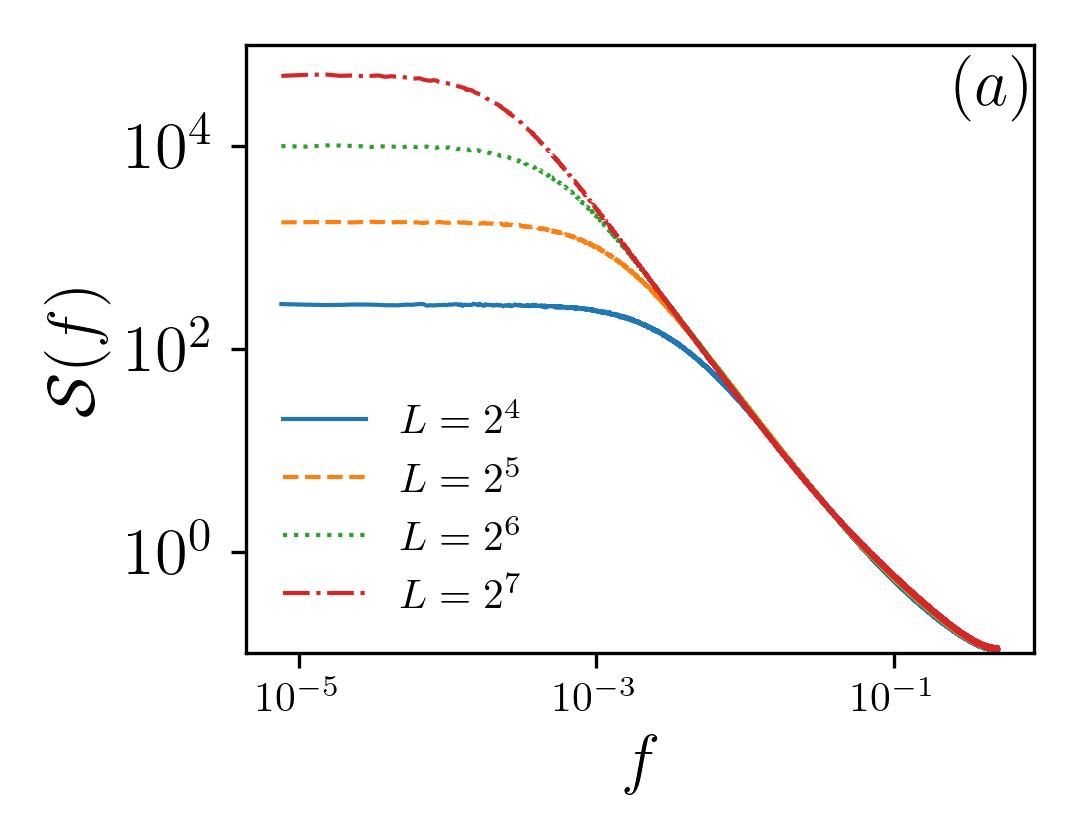}
    \includegraphics[scale=0.45]{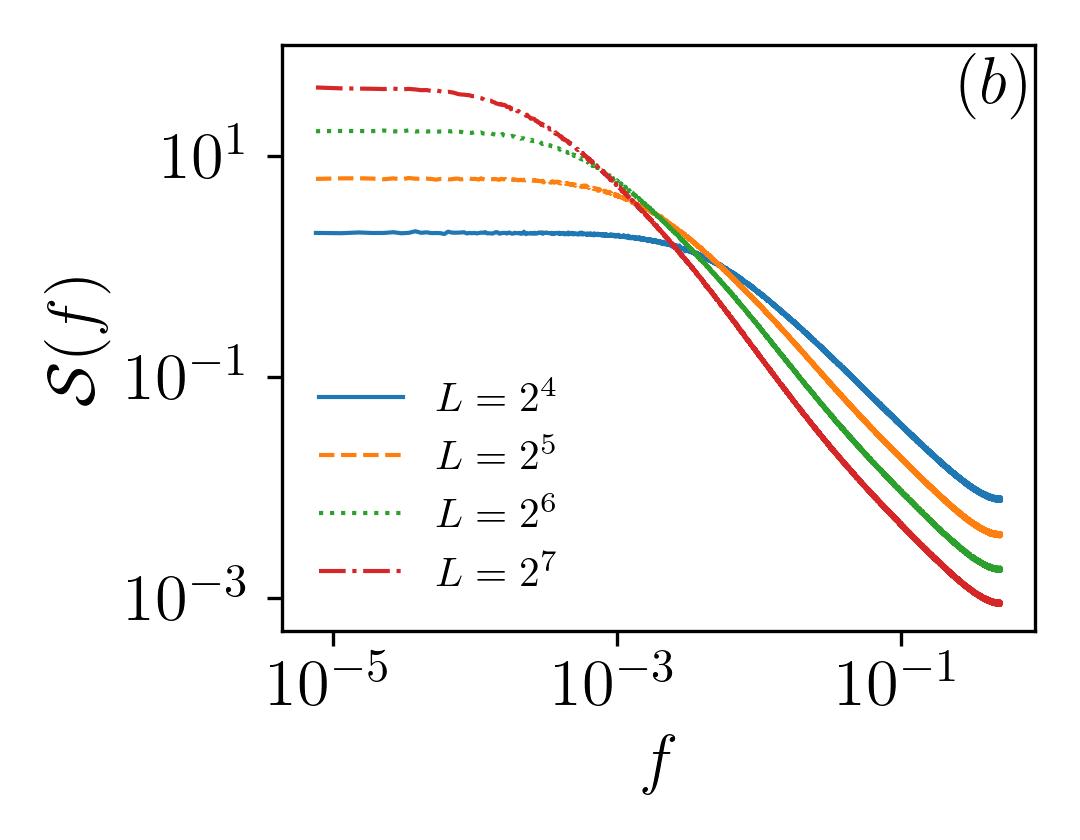} \\
    \includegraphics[scale=0.45]{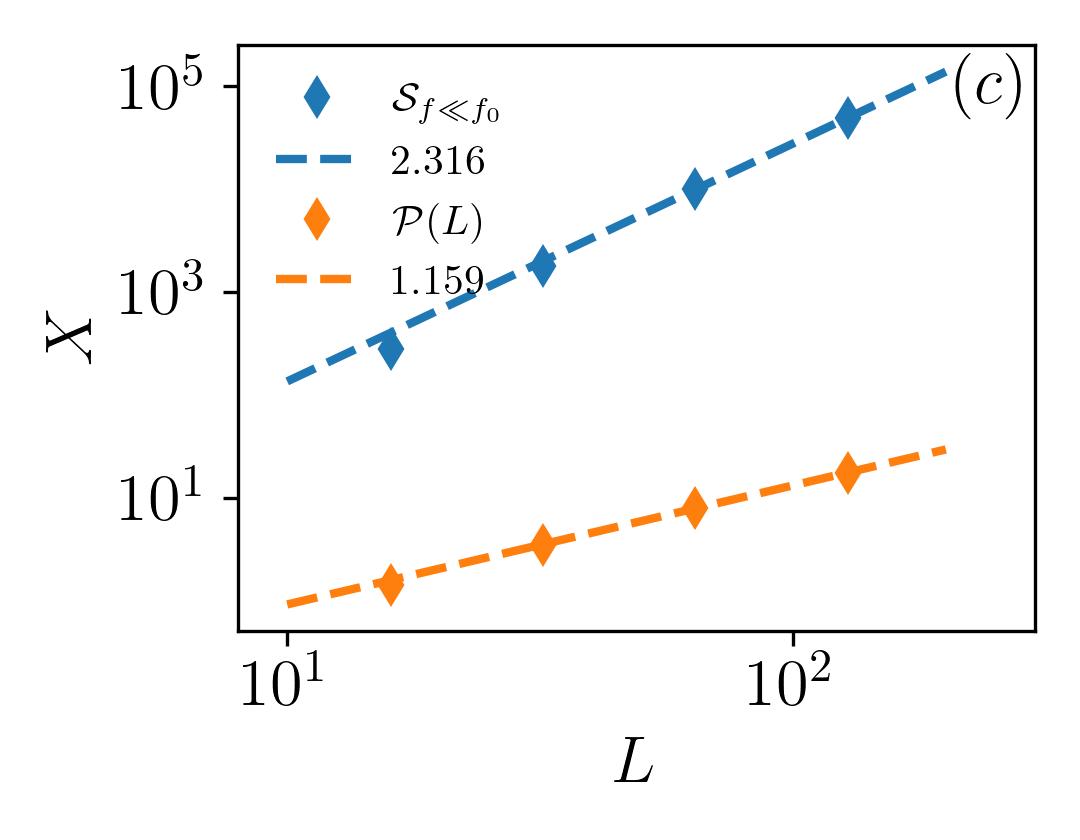} 
    \includegraphics[scale=0.45]{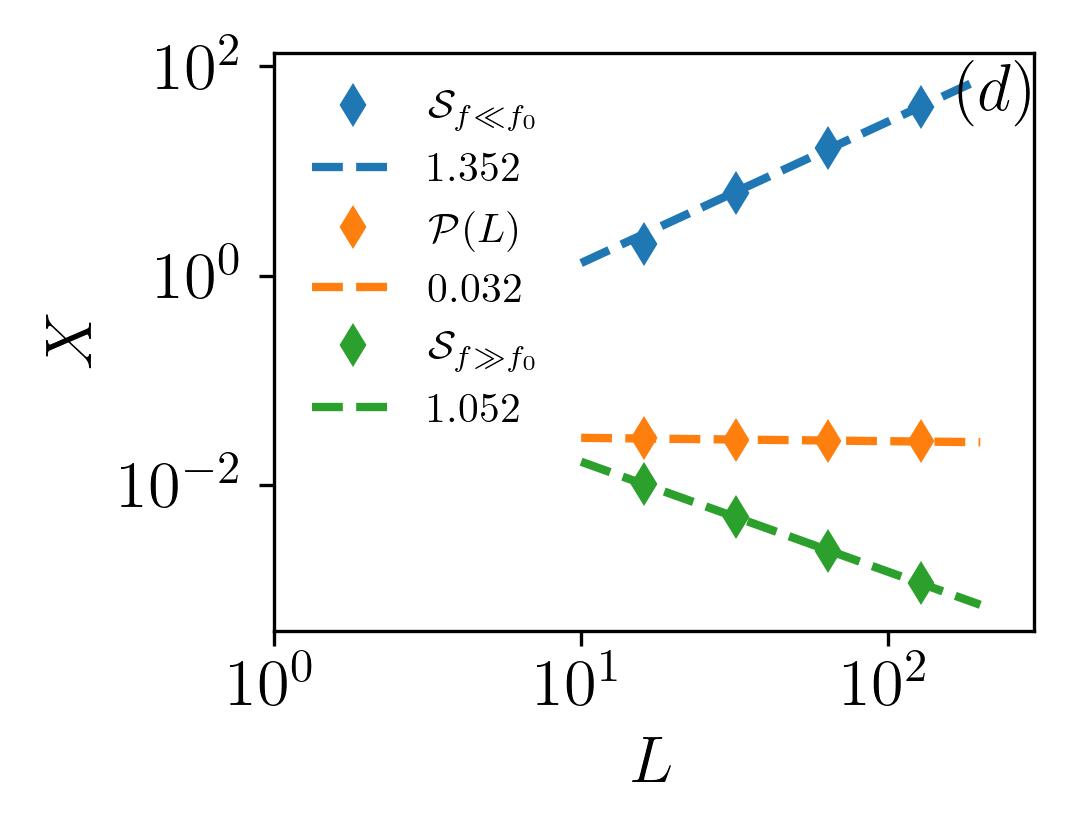} \\
    \includegraphics[scale=0.45]{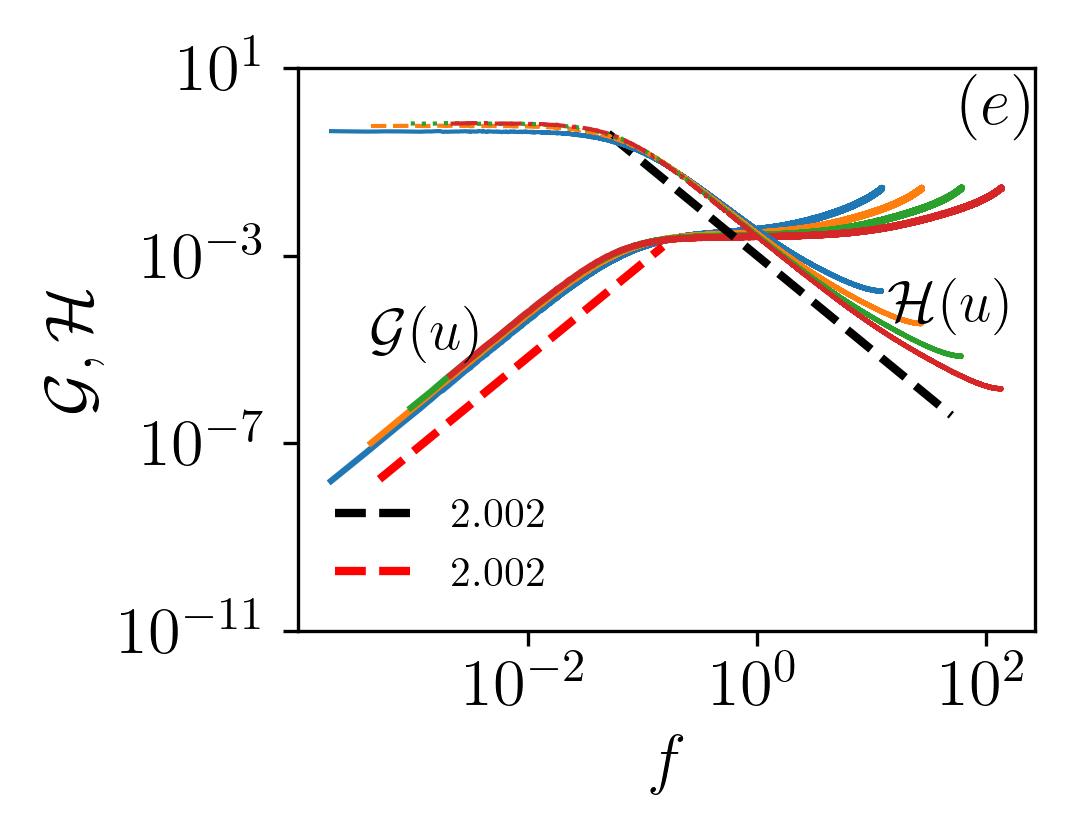}
    \includegraphics[scale=0.45]{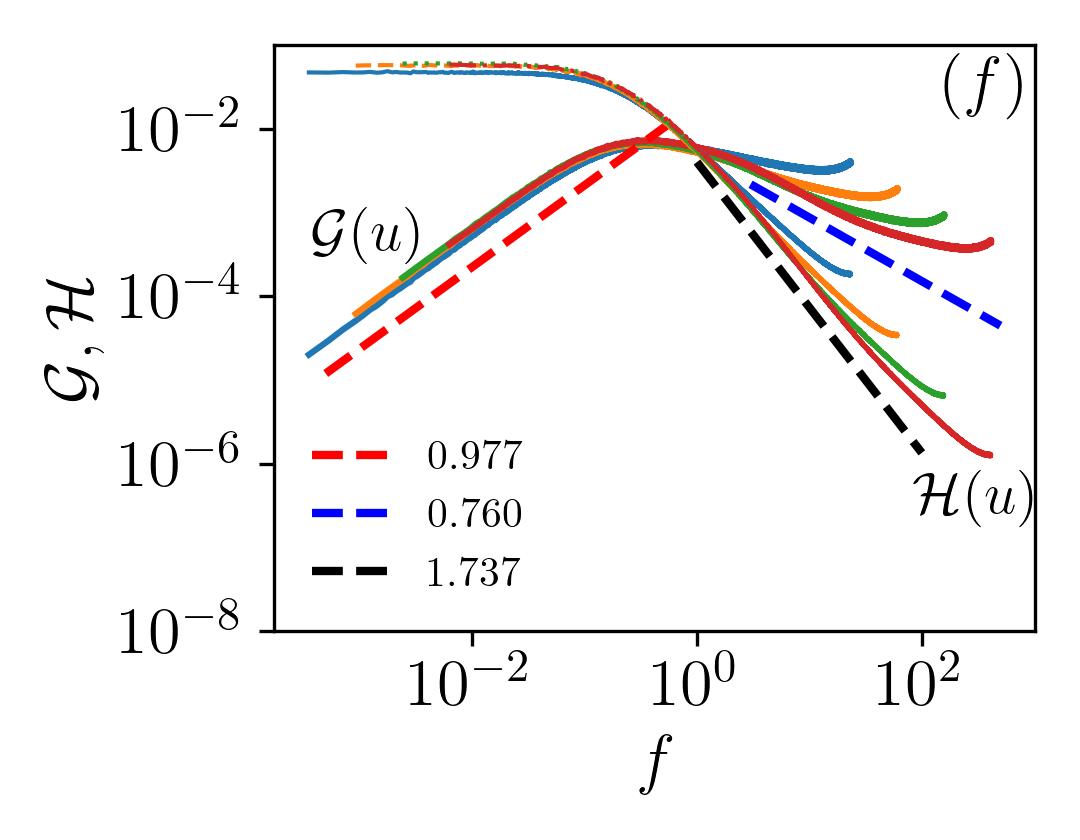}
    \caption{ {\bf For model~C}: Left panel -- {\bf (a)} The power spectra for global fitness fluctuations for system size $L=2^4, 2^5, 2^6, 2^7$; {\bf (c)} The power at a fixed frequency below and above the cut-off $f_0$ including the total power $\mathcal{P}(L)$, along with the best-fit curves. The estimated exponents are $a = 2.32$ and $c = 1.16$. {\bf (e)} The scaling functions for the global fitness power spectra.  Here, $b=0$ which implies  $\lambda = a-c = 1.16$ and $\alpha = a/\lambda = 2.00$. Right panel: Same as left panel, but for local fitness fluctuations. Here, the estimated exponents are $a=1.35,~b=1.05,$ and $c=-0.03$ which imply $\lambda = 2.40$ and $\alpha=(a+b)/\lambda =1.74$.}~\label{Fig-ModelC}
\end{figure}

\begin{table}[htb]
\caption{The critical exponents characterizing the PSD for local $\xi(t)$ and global fluctuations $\eta(t)$ for different variants of the model. The critical exponents $\{a,b,c \}$ are determined from the fitting of straight lines on log-log plots [cf. Fig.~\ref{Fig-Model1}--\ref{Fig-ModelC}(c-d)].}~\label{Tab-CriExp1}%
\centering
\renewcommand{\arraystretch}{1.2} 
    \begin{tabular}{|p{1.2cm}<{\raggedright}|*{5}{>{\centering\arraybackslash}p{0.04\textwidth}|}>{\centering\arraybackslash}p{0.03\textwidth}|*{2}{>{\centering\arraybackslash}p{0.05\textwidth}|}}
    \hline
        {\bf Model}    &   $x(t)$  &   $a$   &   $b$   &   $c$   & $\lambda$ &  $\alpha$     &   $a/\lambda$      &   $b/\lambda$ \\
    \hline
        \multirow{2}{10em}{Model 1}  
            &   $\xi(t)$     & $2.34$   & $1.00$     & $-0.11$    & $2.45$    & $1.36$   & $0.96$  & $0.41$  \\
            &   $\eta(t)$    & $3.36$   & $0.0$     & $1.10$     & $2.26$    & $1.48$   & $1.48$   &  $0.0$ \\
    \hline 
        \multirow{2}{10em}{Model A}  
            &   $\xi(t)$     & $1.62$  & $0.96$     & $-0.1$  & $1.70$     & $1.50$  & $0.94$  &  $0.56$ \\
            &   $\eta(t)$    & $2.54$   & $0.0$     & $1.05$  & $1.49$     & $1.70$  & $1.70$  &  $0.0$  \\
    \hline 
        \multirow{2}{10em}{Model B}  
            &   $\xi(t)$     & $2.19$   & $1.00$     & $-0.13$  & $2.30$  & $1.38$   & $0.95$  & $0.43$  \\
            &   $\eta(t)$    & $3.18$   & $0.0$     & $1.07$   & $2.12$   & $1.50$  & $1.50$  &  $0.0$  \\
    \hline 
        \multirow{2}{10em}{Model C}
            &   $\xi(t)$     &  $1.35$  & $1.05$ & $-0.03$ & $1.38$ & $1.74$ &$0.98$  & $0.76$   \\
            &   $\eta(t)$    &  $2.32$  & $0.0$ & $1.16$ & $2.00$  & $1.16$  & $2.00$  & $0.0$  \\
    \hline
    \end{tabular}

\end{table}

Figs.~\ref{Fig-Model1} correspond to the PSD analysis for the original barycentric BS model, termed as model 1. 
As shown in Figs.~\ref{Fig-Model1}(a) \& (b), the PSD $\mathcal{S}(f)$ as a function of frequency $f$ shows two distinct regimes. For $f \ll f_0$, the PSD is independent of frequency but scales with system size as $\mathcal{S}(f)\sim L^a$. For $f \gg f_0$, the PSD exhibits a $1/f^{\alpha}$ feature, and the scaling with $L$ changes to $\mathcal{S}(f)\sim 1/(L^{b}f^{\alpha})$. 
Therefore, the PSD can be written a a function of frequency $f$ and system size $L$ as
\beqr 
\mathcal{S}(f,L) \sim \begin{cases}
            L^a,~~~~ & f \ll f_0, \\
            f^{-\alpha}L^{-b}, & f \gg f_0,
            \end{cases} \label{eq-PSD1}
\eqnr
which is a homogeneous function of $f$ and $L$. Introducing the reduced frequency $u=fL^{\lambda}$, the PSD takes the form \cite{Singh_2023}
\beqr 
\mathcal{S}(f,L) \sim \begin{cases}
    1/f^{a/\lambda}\, \mathcal{G}(u), & f \ll f_0, \\
    1/f^{\alpha-b/\lambda}\, \mathcal{G}(u), & f \gg f_0,
    \end{cases} \label{eq-PSD2}
\eqnr 
or equivalently,
\beq 
\mathcal{S}(f,L) \simeq \dfrac{1}{f^{a/\lambda}} \mathcal{G}(u) = L^a \mathcal{H}(u),
\eqn
where the scaling functions are
\beqr 
\mathcal{G}(u) \sim \begin{cases}
        u^{a/\lambda}, & u \ll 1, \\
        1/u^{b/\lambda}, & u \gg 1,
        \end{cases} \label{Eq:Gu}
\eqnr 
and 
\beqr
\mathcal{H}(u) \sim \begin{cases}
        \text{constant}, & u \ll 1, \\
        1/u^{(a+b)/\lambda}, & u \gg 1.
        \end{cases} \label{Eq:Hu}
\eqnr 
However, the relation between the exponents can be expressed as 
\beq 
\alpha = \dfrac{a+b}{a-c} = \dfrac{a+b}{\lambda}. \label{Eq-al}
\eqn 

For the global fitness case, we observe $b=0$ [cf. Figs.~\ref{Fig-Model1}(a)]. The total power of the signal also scales as $P(L) \sim \int df \mathcal{S}(f,L) \sim  L^{a-\lambda}$. Figs.~\ref{Fig-Model1}(c) $\&$ (d) show the system size scaling of the power in the frequency regime $f\ll f_0$ and $f\gg f_0$, including the total power $P(L)$. These qualities $X \in \{ S_{f\ll f_0}, S_{f\gg f_0}, P(L) \}$ scale with system size $L$. Thus, we estimate the critical exponents using the best-fit.

Figure~\ref{Fig-Model1}(e)-(f) shows the scaling functions $\mathcal{G}(u)$ and $\mathcal{H}(u)$ for the PSD of the total fitness $\eta(t)$ and local fitness fluctuation signal $\xi(t)$, respectively. The PSD curves for different $L$ collapse onto single-valued scaling functions, confirming the homogeneity of $\mathcal{S}(f,L)$. For $u \ll 1$, $\mathcal{G}(u) \sim u^{a/\lambda}$ and $\mathcal{H}(u)$ tends to a constant, whereas for $u \gg 1$, $\mathcal{G}(u) \sim 1/u^{b/\lambda}$ and $\mathcal{H}(u)\sim 1/u^{\alpha}$, in agreement with Eqs.~\eqref{Eq:Gu}--\eqref{Eq:Hu}.  

The scaling collapse provides a good estimate of the critical exponents $\alpha$ and $\lambda$. These two exponents determine the universality class, $\alpha$ characterizes the slope of the PSD in the high-frequency regime, while $\lambda$ controls the crossover frequency $f_0 \sim L^{-\lambda}$. They are obtained independently from the low-frequency scaling and from the system-size dependence of the total power $P(L)$. Moreover, the local PSD is independent of the lattice site $i$, showing that the local fitness noise remains spatially uncorrelated, as in the classical BS model~\cite{Singh_2023}.

We performed the same PSD and finite-size scaling analysis for the variants (Models~A, B, and C) to test the robustness of SOC under modified interaction rules. 
Despite differences in interaction rule mechanism, the qualitative features remain consistent: the PSD exhibits two regimes with a flat spectrum at low frequencies and a $1/f^{\alpha}$ decay at higher frequencies, followed by good data collapse under finite-size scaling [cf. Figs.~\ref{Fig-ModelA}-\ref{Fig-ModelC}(a-d)]. 
Quantitatively, however, the critical exponents $(\alpha,\lambda)$ differ across models, as summarized in Table~\ref{Tab-CriExp1}, indicating the modification in the correlations to the choice of interaction rule. 
In particular, stochastic or long-range updates tend to shift the exponents away from those of the original barycentric BS model, suggesting possible crossover effects. 
Nevertheless, the persistence of $1/f^{\alpha}$ scaling and data collapse across all variants demonstrates that SOC is a robust emergent property of the model, though its scaling characteristics are non-universal and depend on the specific interaction mechanism. The scaling collapse of $\mathcal{G}(u)$ and $\mathcal{H}(u)$ remains robust, confirming that the PSD retains homogeneous scaling despite the altered interaction rule [cf. Figs.~\ref{Fig-ModelA}-\ref{Fig-ModelC}(e-f)].

\subsection{The cover time and scaling}~\label{sec-corr}

\begin{figure}[htb]
    \centering
    \includegraphics[scale=0.38]{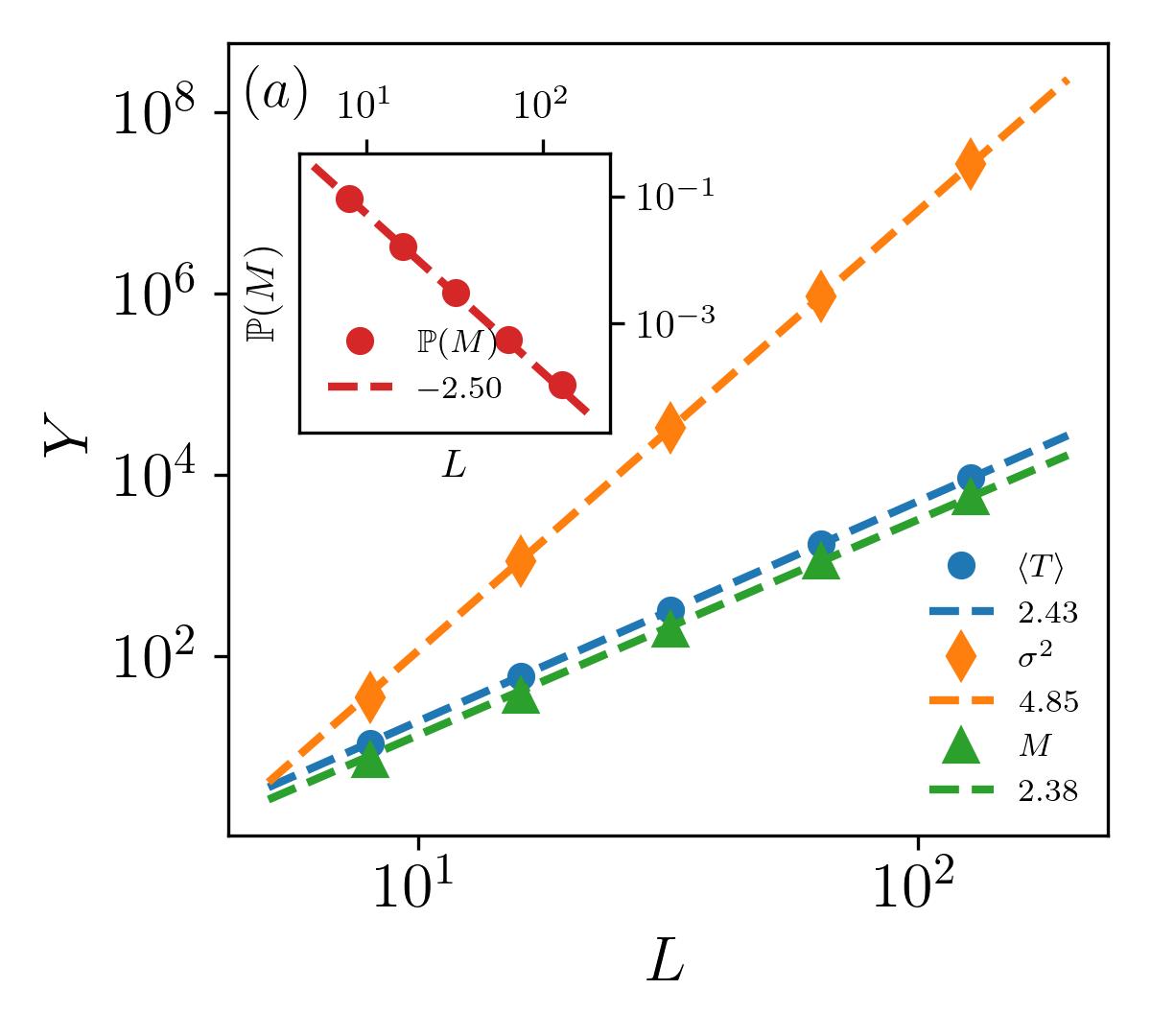}
    \includegraphics[scale=0.5]{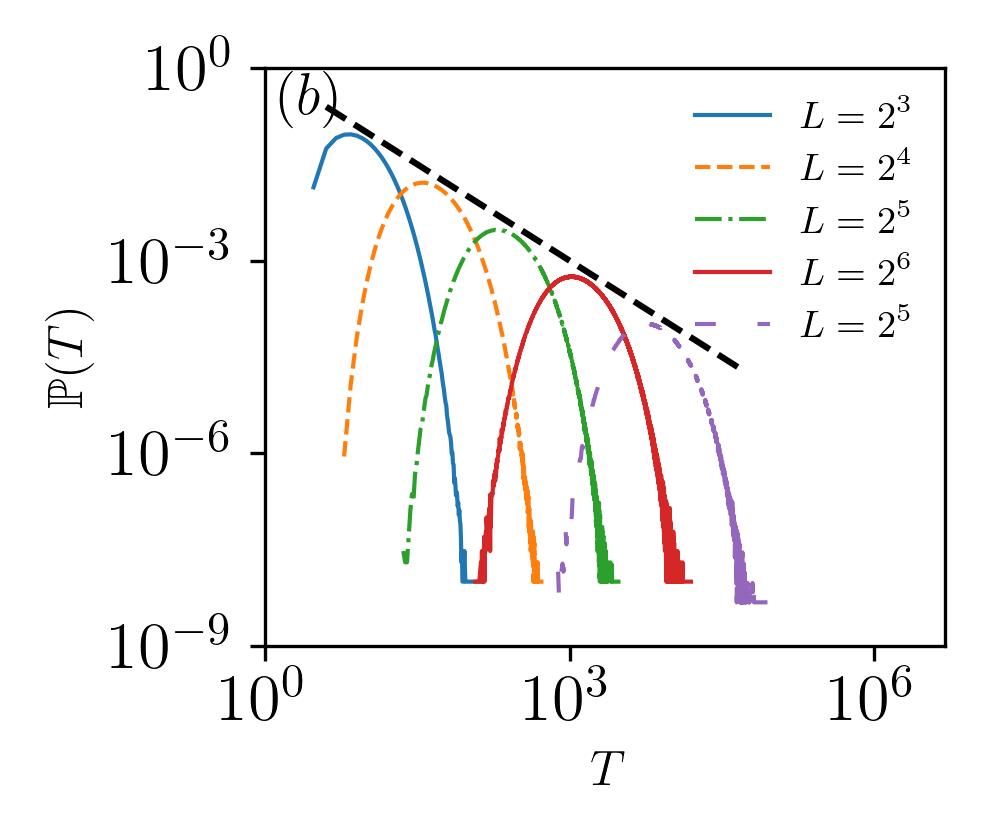}
    \caption{Left panel: The system size scaling for mean $\langle T \rangle$, variance $\sigma^2_T$ and mode $M_T$ of the cover time $T$. In the inset, we plot the probability for the mode ${P}(M)$. The straight line represents the best-fit, along with the estimated critical exponents. Here $Y \in \{ \langle T \rangle, \sigma^2_T, M_T \}$. Right panel: The probability distribution of cover time $T$ in the case of the barycentric BS model. The most probable value ${P}(M)$ decreases with the system size $L$ as $P(M) \sim L^{-\delta_4} \sim T^{-\delta_4/\delta_3} \sim T^{-1}$, since $\delta_3=\delta_4$ [cf. Eq.~\eqref{eq-delta}]. The dashed line provides the confirmation numerically, as it shows the slope $-1$.}~\label{Fig-T1}
\end{figure}

To study the statistical aspects of cover time, first we examine the system size dependence of various statistical characteristics with different system sizes $L=2^3, 2^4, ..., 2^7.$ 
This cover time is a discrete random variable. We use Monte Carlo simulation to simulate the cover time in the barycentric BS model. We collect $N=10^8$ statistically independent samples of cover time after discarding $10^6$ time-steps.
In the Fig.~\ref{Fig-T1}(a), the mean ($\la T \ra $) and variance $(\sigma_T^2)$ of the cover time shows system size scaling with $L$ as
\beqr  
\la T \ra  &=& \dfrac{1}{N} \sum_{i=1}^N T_i \sim L^{\delta_1}~\label{eq-mean} \\ 
\sigma_T^2 &=& \sum_{i=1}^N (T_i - \la T \ra)^2 \sim L^{2\delta_2}~\label{eq-var}
\eqnr
while the mode $(M)$ of the cover varies as 
\beq 
M_T \sim L^{\delta_3}~\label{eq-mode}
\eqn 
and the probability of the cover time at $T=M$ shows the power-law behavior with system size $L$. Thus, 
\beq 
{P}(M) \sim L^{-\delta_4}~\label{eq-peak}
\eqn 
To understand the scaling behavior, we introduce a scaling variable
\beq 
v = \dfrac{T-M}{\sigma_T} = \dfrac{\Delta T}{\sigma_T}~\label{u}
\eqn 
This variable rescales the peak of the probability distribution at $v=0$. Thus, we expect a scaling function of the form
\beq 
{F}(v) = g \dfrac{{P}(x)}{{P}(M)}~\label{scal-fun}
\eqn 
For a normalized probability distribution, we have $\int {P}(x) dx=1$. Thus, plugging Eqs.~\eqref{eq-mean}-\eqref{u} into Eq.~\eqref{scal-fun}, we obtain
\beq 
\int {P}(x)dx \sim L^{\delta_2 - \delta_4} = 1 \nonumber
\eqn 
where the normalization condition suggests $\delta_2 = \delta_4$. Similarly, the shifted extreme activity
\beq 
\la \Delta T \ra \sim L^{2\delta_2 - \delta_4} \sim L^{\delta_1}
\eqn 
implies that $2\delta_2 - \delta_4 = \delta_1$ yielding $\delta_1 = \delta_2$. Since $u$ is independent of the system size, we have $M \sim L^{\delta_3 = \delta_4}$. Thus, 
\beq 
\delta_1 = \delta_2 = \delta_3 = \delta_4 = \delta.~\label{eq-delta}
\eqn 

Then, from Eq.~\eqref{scal-fun}, the probability distribution function can be written as,
\beq 
{P}(x) = \dfrac{1}{L^\delta} {F} \left( u \right) = \dfrac{1}{T}{G} \left( u \right).~\label{Fig-Scale}
\eqn 

\begin{figure}[htb]
    \centering
    \includegraphics[scale=0.45]{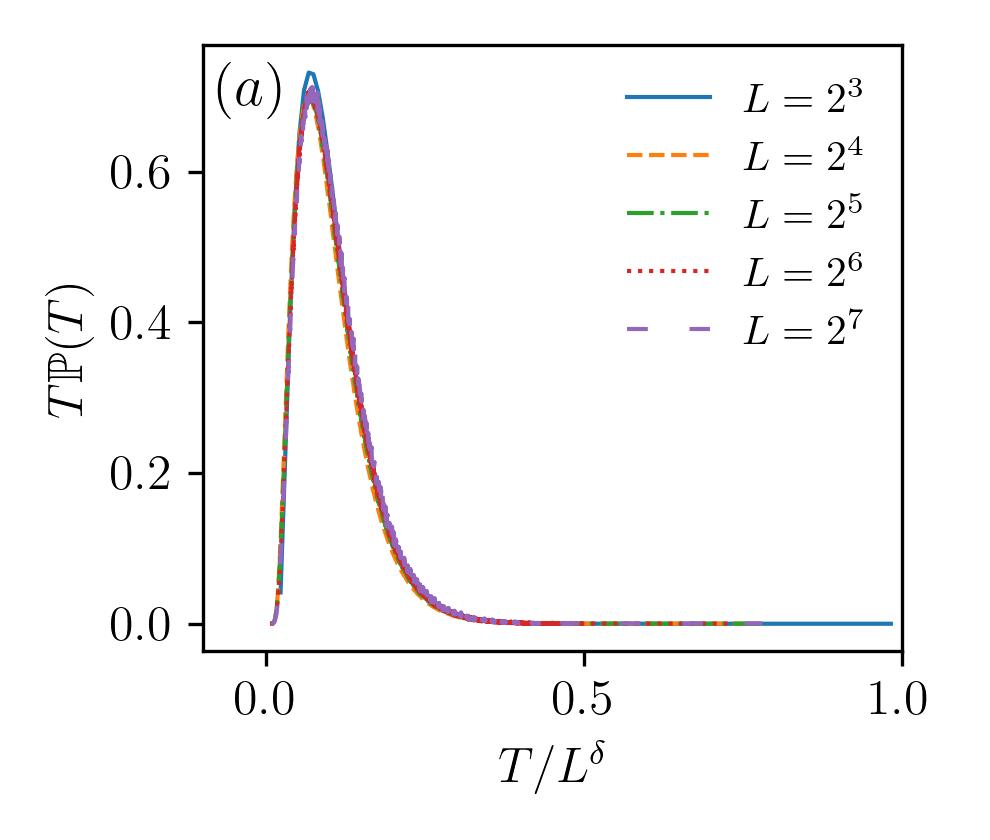}
    \includegraphics[scale=0.45]{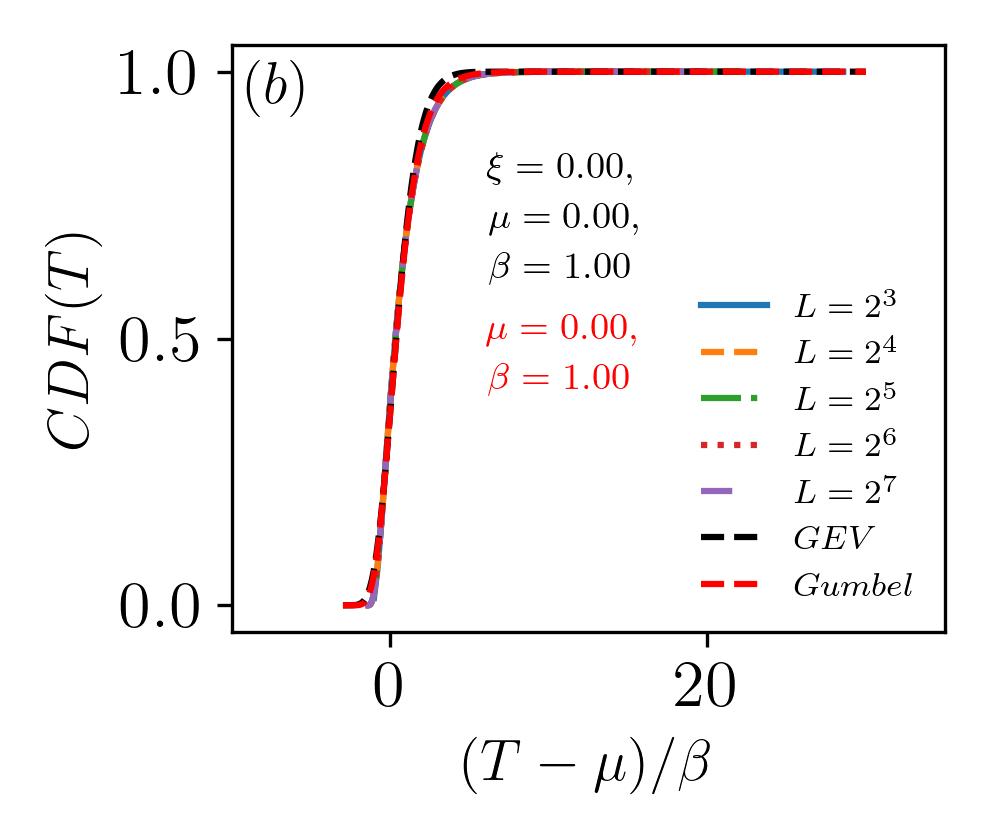}
    \caption{ (a) In the barycentric BS model, the data collapse curve for the PDF and (b) the CDF of cover time, corresponds to Fig.~\ref{Fig-T1}(b).}~\label{Fig-Coll}
\end{figure}

The plot between $T{P}(T)$ and $T/L^\delta$ can provide the scaling function ${F}(u)$ [cf. Fig.~\ref{Fig-Coll}(a)].

We examine the statistical aspects of cover time and report the summary of the critical exponents for different variants of the model in Table~\ref{Tab-CriExp2}. As shown in Table~\ref{Tab-CriExp2}, the estimated exponents $\delta$ are in good agreement with the cover time exponents $\lambda$, studied in Sec.~\ref{sec-fitness}. To substantiate, we consider $CDF(T) = \int {P} (T) dT$. Fig~\ref{Fig-Coll}(a) reveals the good data collapse of {\it CDF}. We fit {\it CDF} with the GEV and Gumbel distributions using the Levenberg-Marquardt algorithm (LMA) method~\cite{LMA1, LMA2}. The CDF fit $\xi=0.00$ implies the presence of the Gumbel distribution~\cite{Coles_2001, Haan_2006}. The fitting parameters of GEV for different variants of the model are shown in Table~\ref{Tab-GEVPar}, which shows consistency with $\xi \sim 0.0$.

\begin{table}[htb]
\caption{Critical exponents for correlation time for different variants of the barycentric BS model.}
\label{Tab-CriExp2}
\centering
\renewcommand{\arraystretch}{1} 
\begin{tabular}{|p{0.08\textwidth}|p{0.06\textwidth}|p{0.06\textwidth}|p{0.06\textwidth}|p{0.06\textwidth}|p{0.06\textwidth}|}
\hline 
\textbf{Model} & $\delta_1$ & $\delta_2$ & $\delta_3$ & $\delta_4$ & $\lambda$ \\
\hline 
Model 1 & 2.43 & 2.43 & 2.38 & 2.50 & 2.26 \\
\hline 
Model A & 1.74 & 1.70 & 1.81 & 1.70 & 1.49 \\
\hline 
Model B & 2.41 & 2.42 & 2.42 & 2.33 & 2.12 \\
\hline
Model C & 1.74 & 1.75 & 1.78 & 1.84 & 1.70 \\
\hline
\end{tabular}
\end{table}

\begin{table}[htb]
\caption{The fitted parameters, describing the scaling function for the probability distribution of correlation time for different variants of the model. In all the cases, the goodness of fit is $R^2 > 0.99$.}~\label{Tab-GEVPar}
\centering
\renewcommand{\arraystretch}{1.2} 
\begin{tabular}{|p{0.08\textwidth}|p{0.07\textwidth}|p{0.07\textwidth}|p{0.07\textwidth}|p{0.07\textwidth}|p{0.07\textwidth}|}
\hline
\multirow{2}{*}{\bf Model} & \multicolumn{2}{c|}{\bf Gumbel} & \multicolumn{3}{c|}{\bf GEV} \\ \cline{2-6}
                       & $\mu$ & $\beta$ & $\mu$ & $\beta$ & $\xi$ \\ \hline
Model 1   & 0.0 & 0.99 & 0.0 & 0.98 & 0.10 \\ \hline
Model A  & -0.02(1) & 1.0 & 0.0 & 0.99 & 0.10 \\ \hline
Model B  & -0.02(1) & 0.99 & 0.0 & 0.98 & 0.09 \\ \hline
Model C  & -0.02(1) & 0.99 & 0.0 & 0.99 & 0.09 \\ \hline
\end{tabular}
\end{table}

\section{Conclusion}~\label{sec-conclusion}

In summary, we have studied the one-dimensional barycentric BS model and their variants. The model demonstrates self-organized criticality, and as expected in such systems, we observe the emergence of long-range space-time correlations. The trajectory of the least conformist site exhibits L\'evy flight behavior, leading to a fractal structure in the space-time plane. We examined fluctuations in fitness using power spectra for different system sizes. The finite-size scaling (FSS) analysis yields the scaling functions and critical exponents through data collapse. The fitness fluctuations follow the $1/f^{\alpha}$ form with the spectral exponent $\alpha \sim 1.4$. The cutoff frequency varies as $f_0 \sim L^{-\lambda}$ with $\lambda = 2.45$ for local fitness fluctuation and $\lambda=2.26$ for global fitness fluctuation. The global fitness shows uncorrelated behavior in the absence of explicit interaction.  Although the barycentric BS model and the classical BS model do not belong to the same universality class~\cite{Kennerberg_2021}, both display $1/f^{\alpha}$ behavior in the non-trivial frequency regime. In this study, we have presented two different routes to estimate the critical exponent $\lambda$ numerically. The FSS and scaling function analysis provide insights into the $1/f^{\alpha}$ behavior as well as correlation times. 

Beyond the classical case, we analyzed several variants of the barycentric BS model to test the robustness of SOC under modified interaction rules. While the qualitative signatures of SOC, such as $1/f^{\alpha}$ scaling and finite-size data collapse, persist across all variants, the critical exponents $(\alpha,\lambda)$ exhibit model-dependent shifts, as summarized in Table~\ref{Tab-CriExp1}. These variations highlight the dependency of critical exponents to the nature of the interaction mechanism, with stochastic or long-range updates leading to noticeable crossover effects. Nevertheless, the persistence of fractal trajectories, L\'evy flight statistics, and homogeneous scaling across all models confirms that SOC remains a robust emergent property of the barycentric BS framework, though its quantitative characteristics are non-universal and depend on the specific interaction rule.

We also examined the fitness cover time, defined as the duration required to update (extinction or mutation) the fitness of all species in the entire system. Monte Carlo simulations suggest that the statistical properties of the cover time follow a power-law distribution as a function of system size with the same critical exponent. The peak of the probability distribution of the cover time also scales as ${P}(T) \sim T^{-1}$. Employing FSS with generalized extreme value (GEV) theory, we proposed the corresponding scaling function. The data collapse shows good agreement with the Gumbel distribution. The numerically estimated cover time exponents are consistent across different methods. Notably, the universal behavior is reflected in the fact that the correlation time distribution collapses onto the Gumbel distribution. 

Taken together, our findings demonstrate that SOC in the barycentric BS framework is both resilient and sensitive: resilient in that scale-free behavior, fractal dynamics, and L\'evy flights persist across variants, yet sensitive in that the quantitative values of critical exponents depend on the precise interaction rules. An interesting direction for future work is to extend this study to higher-dimensional barycentric BS models and explore whether the observed robustness and non-universality persist.

\section{acknowledgments}
AQ greatly acknowledges support from the INSPIRE Fellowship (DST/INSPIRE Fellowship/IF180689), under the Department of Science and Technology, Government of India. 
Computational resources were provided by the PARAM Rudra Supercomputing Facility at the Inter-University Accelerator Centre (IUAC), New Delhi, under the National Supercomputing Mission (NSM), Government of India.

\end{document}